\documentclass[fleqn,11pt,letterpaper]{article}
\usepackage{amsfonts}
\usepackage{amssymb}
\usepackage{amsmath}
\usepackage{amsthm}
\usepackage{enumerate}
\usepackage{multirow}
\usepackage{float}
\usepackage{graphicx}
\usepackage{epstopdf}
\usepackage{color}
\usepackage[margin=2.57cm]{geometry}
\definecolor{darkred}{rgb}{0.5,0.2,0.2}
\usepackage[pdftitle={SRCM},pdfauthor={Mikkel Bennedsen, Eric Hillebrand, Siem Jan Koopman},colorlinks=TRUE,allcolors=darkred]{hyperref}
\usepackage{booktabs}
\usepackage{pdflscape}
\usepackage[round]{natbib}

\usepackage{tikz}
\usetikzlibrary{arrows.meta,
                chains,
                positioning,
                shapes.geometric 
                }

\usepackage{tabularx}

\usepackage{float}
\floatstyle{plaintop}
\restylefloat{table}

\restylefloat{table}

\theoremstyle{plain}

\newcommand{\R}{\mathbb{R}}

\def\bi{\begin{itemize}}
\def\ei{\end{itemize}}

\allowdisplaybreaks
\graphicspath{{Figures/}}
\numberwithin{equation}{section}

\newif\ifi

\title{A Statistical Reduced Complexity Climate Model for \\ Probabilistic Analyses and Projections}

\author{Mikkel Bennedsen\thanks{
Department of Economics and Business Economics and CoRE, 
Aarhus University, 
Fuglesangs All\'e 4,
8210 Aarhus V, Denmark.
E-mail:\
\href{mailto:mbennedsen@econ.au.dk}{\nolinkurl{mbennedsen@econ.au.dk}} 
},
Eric Hillebrand\thanks{
Department of Economics and Business Economics and CoRE, 
Aarhus University, 
Fuglesangs All\'e 4,
8210 Aarhus V, Denmark.
E-mail:\
\href{mailto:ehillebrand@econ.au.dk}{\nolinkurl{ehillebrand@econ.au.dk}} 
},
Siem Jan Koopman\thanks{
Department of Econometrics, 
School of Business and Economics,
Vrije Universiteit Amsterdam, 
De Boelelaan 1105,
1081 HV Amsterdam, The Netherlands,
E-mail:\
\href{mailto:s.j.koopman@vu.nl}{\nolinkurl{s.j.koopman@vu.nl}} 
}}

\begin{document} 
\maketitle

\begin{abstract}
We propose a new
statistical
reduced complexity climate model.
The centerpiece of the model consists of a set of physical equations for the global climate system
which we show how to cast in non-linear state space form.
The parameters in the model are estimated using the method of maximum likelihood with
the likelihood function being evaluated by the extended Kalman filter.
Our statistical framework is based on well-established methodology and is computationally feasible.
In an empirical analysis, we estimate the parameters for a data set comprising the period $1959$--$2022$.
A likelihood ratio test sheds light on the most appropriate equation for converting the level of atmospheric
concentration of carbon dioxide into radiative forcing.
Using the estimated model, and different future paths of
greenhouse gas emissions, we project global mean surface
temperature until the year $2100$.
Our results illustrate the potential of combining statistical modelling with physical insights to arrive at
rigorous statistical analyses of the climate system.
\end{abstract}

\newpage
%
%
%
%

\section{Introduction}\label{sec:intro}
Climate models play an important role in our understanding of the past, present, and future climate. Most notably, the reports on the ``physical science basis'' for global warming, compiled by the Intergovernmental Panel on Climate Change (IPCC), rely heavily on output from large-scale Earth system models \citep{IPCC2007_4th,IPCC2013_5th,IPCC2021_6th}. However, since these large-scale models require enormous amounts of computer power to run even a single instance, they are not suited for creating large ensembles of climate system variables, which could be useful for, e.g., probabilistic analysis and/or for studying the effects of a large number of different scenarios for future greenhouse gas emissions.  In contrast to large-scale Earth system models, reduced complexity climate models (RCMs) are simpler climate models designed to run on desktop computers. The lower computational requirements of  RCMs make them eminently suitable for running large ensembles and investigating the impact of a wide range of different greenhouse gas scenarios, as well as for probabilistic analysis, which would be infeasible using large scale models \citep[][]{RCMIP2020,RCMIP2021}. RCMs are by now recognized as important tools for understanding the climate system. For instance, the ``MAGICC'' RCM \citep{MAGICC} is instrumental in probabilistic analyses of the climate system \citep{Meinshausen2009} as well as in the construction of the influential Representative Concentration Pathway (RCP) scenarios \citep{Meinshausen2011}. Both MAGICC and the ``FaIR'' RCM \citep{FAIR} are extensively used in the 6'th assessment report from the IPCC \citep[AR6;][see, e.g., Chapter 7]{IPCC2021_6th}.

We consider a statistically-based RCM where the entire chain from emissions to temperatures
is formulated as a statistical model with stochastic representations of variables and their dependencies implied by the underlying physics.
This approach is in contrast to a physically-based RCM which is formulated as a deterministic model
using equations reflecting the underlying physics.
As a consequence, physically-based RCMs are not accounting for stochastic variation and
uncertainty in the modeling framework, and hence the statistical properties of the calibrated/estimated model parameters are unknown.
Furthermore, the informative and insightful ``probabilistic'' projections are not necessarily based on
justified statistical foundations.
Most RCMs that model the entire chain from emissions to temperatures are physically-based \citep{FAIR_calibrate}.\footnote{Notably, the most recent version of the FaIR model \citep{FAIR_calibrate} treats the energy balance module statistically using the framework suggested in \cite{Cummins2020} and the remaining modules  deterministically, putting this model somewhere in-between being statistically-based and physically-based.}

In this paper, we propose and develop a fully statistical reduced complexity climate model (Stat-RCM) that models the entire chain from emissions to temperatures.
We assume that the climate system consists of various latent processes --- such as the atmospheric
concentration of carbon dioxide (CO$_2$) and the mean temperature at the Earth's surface --- and that
historical data are noisy observations of these variables.
This framework allows us to cast the climate model in a state space system,
where the state equations represent the latent ``true'' climate processes and
the measurement equations express observations as noisy measurements of these latent processes.
We then specify the transition equations of the state variables using functional forms which are
chosen to be compatible with the physics underlying the climate system.
The stochastic error processes in the state equation represent the difference between the
latent ``true'' climate system and our functional representation of it.  

Our proposed Stat-RCM treats the  physical processes of the climate system at a lower resolution compared to physically-based RCMs, such as MAGICC \citep{MAGICC}, FaIR \citep{FAIR}, and Hector \citep{HECTOR}. The simpler physical representation of the climate system employed by the Stat-RCM is necessitated by statistical identification, i.e., the condition that the objective function, usually the log-likelihood, as a function of the parameters of the model has a unique extremum. Identification allows for the invocation of standard statistical results, such as a law of large numbers and a central limit theorem. The former ensures that the estimator of the parameter vector is consistent and the latter ensures that the estimator is asymptotically normally distributed, which allows for assessment of parameter uncertainty.  
We use our framework in this study to (i) estimate the parameter vector of the model via the
maximum likelihood method using historical data from the period $1959$--$2022$;
(ii) obtain standard errors for the estimated parameters;
(iii) investigate  the finite sample properties of the estimation method via Monte Carlo simulations; (iv) conduct model selection on the parts of the model that are not informed by physical first-principles; (v) extract estimates of the underlying ``true'' climate state variables; (vi) perform statistical validation of the model by showing that it can simulate climate variables that are compatible with the historical data record; and (vii) construct probabilistic projections of the climate system, conditional on a scenario for future greenhouse gas emissions. Our results  illustrate the potential for combining statistical modelling with physical insight to arrive at rigorous probabilistic and statistical analyses of the climate system. 

The remainder of the paper is structured as follows.
In Section \ref{sec:model}, we formulate the basic model equations of the Stat-RCM, and
in Section \ref{sec:ss} we show how to cast these equations into a non-linear state space system.
Section \ref{sec:sims} presents the results from a Monte Carlo simulation study that aims to investigate the
finite sample properties of the proposed estimation procedure.
Section \ref{sec:empirical} shows and discusses the estimation results for the Stat-RCM based
on historical data from the period $1959$--$2022$.
In Section \ref{sec:projections} we adopt these estimation results and use these to generate projections
of the future climate, conditional on given paths of greenhouse gases and other forcing agents.
Section \ref{sec:concl} concludes. An appendix contains  supplementary empirical results  and details on the implementation of the employed state space methods.

\section{Stat-RCM: A new statistical reduced complexity climate model}\label{sec:model}

An overview of  Stat-RCM is given in Figure \ref{fig:overview}. The figure shows how anthropogenic CO$_2$ emissions from fossil fuels ($E^{FF}$) and land--use change ($E^{LUC}$) are affecting temperatures: first, total CO$_2$ emissions ($E =  E^{FF} + E^{LUC}$) are converted into atmospheric concentrations ($C$) in a carbon cycle module; then atmospheric concentrations are converted into radiative forcing ($F$) using a forcing equation; lastly, an energy balance module converts radiative forcing into changes in surface temperature ($T^m$) and deep ocean temperature ($T^d$). CO$_2$ emissions and the forcing effects of non-CO$_2$ greenhouse gases ($F^{Non-CO2}$) and various natural phenomena such as volcanic eruptions and solar activity  ($F^{Nat}$)  are incorporated into the model as  covariates, i.e. these processes are ``external'' and not determined inside the model. 

The following subsections give further details on the three main modules in Stat-RCM, namely the carbon cycle module, the forcing equation, and the energy balance module. Similar models for the carbon cycle and energy balance modules were considered separately in \cite{BHK2023b} and \cite{BHL2022}, respectively. The Stat-RCM extends these models by connecting them through the forcing equation, thereby modelling the entire chain from emissions to temperatures, and by introducing climate feedbacks into the carbon cycle module.

                    \begin{figure}[ht]
                \centering
        \tikzstyle{block} = [rectangle, draw,
        text width=7em, text centered, rounded corners, minimum height=3em,node distance=2cm]
        \caption{\footnotesize Diagram of Stat-RCM}
                \label{fig:overview}
        \begin{tikzpicture}[node distance = 3cm, auto]
            \footnotesize
            \node [block, text width=2.5cm] (Forcings) {\textbf{Forcing equation} \\$C \rightarrow F$};
            \node [block, left of=Forcings, node distance=4cm] (GCB)[left=-2.1cm of Forcings] {\textbf{Carbon cycle module} \\  $E \rightarrow C$};
            \node [block, above=0.5 cm of GCB, text width=3cm, minimum height=6.0em] (E) {\textbf{External} \\$E = E^{FF} + E^{LUC}$};
            \node [block, right of=Forcings, node distance=4cm, text width=2.5cm] (EBM)[right=-2.1cm of Forcings] {\textbf{Energy balance module} \\ $F \rightarrow T$};
            \node [block, right of=EBM, node distance=4cm] (Out) {\textbf{Output} \\ $T^m$, $T^d$};
            \node [block, above=0.5 cm of EBM, text width=3cm, minimum height=6.0em] (Ex) {\textbf{External} \\ $F^{\textrm{Ex}} = F^{\textrm{Non-CO2}} + F^{\textrm{Natural}}$};
            \draw [-{Stealth[scale=1.2]}] (E) --  (GCB) ;
            \draw [-{Stealth[scale=1.2]}] (GCB) --  (Forcings) ;
            \draw [-{Stealth[scale=1.2]}] (Forcings) -- (EBM);
            \draw [-{Stealth[scale=1.2]}] (EBM) -- (Out);
            \draw [-{Stealth[scale=1.2]}] (Ex) -- (EBM);
        \end{tikzpicture}
        \label{flowchart}
    \end{figure}


\subsection{Carbon cycle module}\label{sec:gcmodule}
The cornerstone of the carbon cycle module is the carbon budget equation,
\begin{align}\label{eq:cb}
\frac{dC_t}{dt} = E^{FF}_t + E_t^{LUC} - S_t^{LND}- S_t^{OCN} ,
\end{align}
where $C_t$ is atmospheric concentration of carbon dioxide (CO$_2$) at time $t$, while $E^{FF}_t$ and $E_t^{LUC}$ are emissions of CO$_2$ at time $t$
from fossil fuel burning and land-use change, respectively, and, $S_t^{OCN}$ and $S_t^{LND}$ are fluxes in the ocean and terrestrial (land) sink, respectively. Atmospheric concentrations, $C_t$, are given in gigatonnes of carbon (GtC), while all terms on the right hand side of Equation \eqref{eq:cb} are given in GtC \emph{per year}.
The carbon budget equation \eqref{eq:cb} captures the principle that Earth's carbon cycle is `closed' in the sense that all CO$_2$ emitted from human activities ($E^{FF}_t + E^{LUC}_t$) must be absorbed by one of the three carbon sinks, namely the atmosphere ($dC_t/dt$), the terrestrial biosphere ($S_t^{LND}$), and the oceanic biosphere ($S_t^{OCN}$);
see \cite{GCB2022short} for an extensive discussion of the  carbon budget.

The emissions time series $E_t^{FF}$ and $E_t^{LUC}$ are considered as covariates (exogenous data), which ``force'' the carbon system. The stock time series $C_t$ as well as the flux time series data $S_t^{LND}$ and $S_t^{OCN}$
are included as part of the dynamic model system and interact with the other variables in the model.
As is common in reduced-form modelling of the carbon cycle, we will assume that the land and ocean fluxes are related to the level of atmospheric concentrations ($C$) and the state of the climate, represented by the global surface temperature above pre-industrial levels ($T^m$). 
In particular, we specify the dynamics of the sinks as
\begin{align*}
S_t^{OCN}& = g_{OCN}(C_t,T_t^m) - g_{OCN}(C_{1750},T_{1750}^m)  + \eta_t^O, \\ 
S_t^{LND} &= g_{LND}(C_t,T_t^m) - g_{LND}(C_{1750},T_{1750}^m) + \eta_t^L, 
\end{align*}
for non-linear functions $g_{OCN}(\cdot,\cdot)$ and $g_{LND}(\cdot,\cdot)$, and where the
terms $\eta_t^O$ and $\eta_t^L$ are zero-mean stochastic processes,
$C_{1750}$ and $T_{1750}^m$ denote the pre-industrial values of $C$ and $T^m$, respectively, with
default values given by $C_{1750} = 591.3060$ GtC and $T_{1750}^m = 0^\circ$C. 
The terms $g_{i}(C_{1750},T_{1750}^m)$, for $i=OCN,LND$, are introduced to capture the assumption
that the climate system was in equilibrium in the pre-industrial era, meaning that
the net flux of CO$_2$ between the atmosphere and the sinks were zero on average,
i.e. $\mathbb{E}[S_{1750}^{OCN}] = \mathbb{E}[S_{1750}^{LND}]  = 0$, where $\mathbb{E}[\cdot]$ denotes the expectation operator.  

Different RCMs consider different functional forms for $g_{OCN}$ and $g_{LND}$.
Over the time period studied in the empirical section below ($1959$--$2022$),
the functions  $g_{OCN}$ and $g_{LND}$ can be taken to be approximately linear in concentrations $C$ \citep[][]{BHK2023b}. We follow the MAGICC RCM \citep[][]{MAGICC} approach and incorporate $T^m$ via exponential functions,
\begin{align*}
g_{OCN}(C_t,T_t^m)  &= b_1 C_t \exp(-c_1 T_t^m), \\
g_{LND}(C_t,T_t^m)  &= b_2  C_t\exp(-c_2 T_t^m), 
\end{align*}
where $b_1,b_2, c_1, c_2 \in \R$ are constant parameters. 
For $b_1,b_2 >0$, the activity of the sinks depends positively on the level of
atmospheric concentrations $C$, and we provide evidence of this below.
For the land sink, this is due to the fertilization effect \citep[e.g.][]{bacastow1979models},
while  for the ocean sink it is due to the fact that the uptake depends on the difference in
partial pressure of CO$_2$ between the atmosphere and the upper ocean \citep[e.g.][]{Joos1996}.
The dependence of the sink uptake  on the level of CO$_2$ concentrations in the atmosphere, is often
referred to as the `carbon cycle feedback'.
Conversely, the dependence of the sink uptakes on the state of the climate, represented by the
global surface temperature $T^m$, is referred to as the `climate feedback' \citep[e.g.][]{Friedlingstein2003,Fung2005,Friedlingstein2015}.

\subsection{Forcing equation}\label{sec:Feq1st}
The forcing equation connects the level of CO$_2$ concentrations (given in GtC) to radiative forcing from CO$_2$ (given in W/m$^2$) above the pre-industrial level. Here, it is specified as
\begin{align}\label{eq:F}
F_{t}^{CO2} = g_{F}(C_t) - g_{F}(C_{1750}) + \eta_{t}^F,
\end{align}
where $g_{F}(\cdot)$ is a non-linear function describing the link between CO$_2$
concentrations and forcing, $\eta_{t}^F$ is a zero-mean stochastic process, and
the term $g_F(C_{1750})$ is introduced to reflect the assumption that the climate system was in equilibrium in the pre-industrial era, i.e. $\mathbb{E}[F_{1750}^{CO2}] = 0$.
In this study, we consider the following functional form for $g_{F}()$
\begin{align}\label{eq:F2}
g_{F}(C_t) = f_1\log \left(C_t +f_2 C_t^2\right)  + f_3\sqrt{C_t},
\end{align}
where $f_1, f_2, f_3  \in \mathbb{R}$ are unknown constant parameters.
This specification for $g_F()$ is general and it nests several functional forms which have been used in the RCM literature previously. Setting $f_1 = f_2 = 0$ will result in the specification used in MAGICC6 
\citep{MAGICC}; setting $f_2 = 0$ will result in the specification used in FAIR \citep{FAIR}; while $f_3 = 0$ yields the form proposed in \cite{Hansen1998}. Below, we will exploit the statistical nature of the model to test which of these functional forms are most appropriate when confronted with historical data.

\subsection{Energy balance module}\label{sec EBM}
For the energy balance module, we consider a two-box energy balance model \citep[][]{gregory2000vertical}
\begin{align*}
H_m \frac{d T_t^{m}}{dt} &= F_t - \lambda T_t^{m} - \gamma( T_t^{m} - T_t^{d}), \\
H_d \frac{d T_t^{d}}{dt} &=  \gamma( T_t^{m} - T_t^{d}), 
\end{align*}
where $H_m$ and $H_d$ are fixed unknown coefficients, 
$T_t^{m}$ and $T_t^{d}$ are time-$t$ temperature anomalies
(with respect to the pre-industrial period) in the atmosphere upper-ocean mixed layer and the
deep ocean layer, respectively,  $F_t$ is time-$t$ radiative forcing, and $\lambda$ and $\gamma$ are fixed unknown coefficients.
The term $F_t = F_t^{CO2} + F_t^{Ex}$ is the total
radiative forcing in the system which we model as the sum of the forcing from CO$_2$ ($F_t^{CO2}$ is modelled within the dynamic system) and forcing  from other sources ($F_t^{Ex}$ represents a set of covariates).
The parameters $H_m$ and $H_d$ (given in $W$ year $m^{-2} K^{-1}$) are the corresponding heat capacities for the two layers, $\lambda$ (given in $W$ $m^{-2} K^{-1}$) is a climate feedback parameter, and $\gamma$ (given in $W$ $m^{-2} K^{-1}$) is the coefficient of heat exchange between the two layers.

In this study, we work with the two-box energy balance model as described above.
However, it is straightforward to expand our setting to a general model with one box or multiple boxes.
For instance, a three-box energy balance model can be considered using related methods to
those considered in \cite{Cummins2020}.

\section{Non-linear state space representation of Stat-RCM}\label{sec:ss}

In this section, we show how to cast the climate model presented in Section \ref{sec:model} in a
non-linear state space system. We do this by using an Euler discretization of the relevant
continuous-time equations.
A state space system consists of two sets of equations: $(i)$ state equations for the latent system dynamic variables, and $(ii)$ measurement equations for the observed data variables.
In Section \ref{sec:state}, we present the state equations
which are obtained from the physical equations above.
In Section \ref{sec:measurement}, we present the measurement equations
which specify the data observations as noisy measurements of the latent system variables. 
In Section \ref{sec:system}, we collect the state and measurement equations into a
state space system, which can then be estimated and analyzed using statistical methods
designed for such systems. We refer to \cite{durbin2012time} for a textbook treatment on the statistical analysis of state space systems. 

In what follows, let $\Delta > 0$ denote the length of a time step. In our applications of the model presented below, we will set $\Delta = 1$ year, but, for generality, we specify the model for a generic time step.

\subsection{State equations}\label{sec:state}

The carbon cycle module equations are discussed in Section \ref{sec:gcmodule} and, after some
substitutions and minor manipulations, we have
\begin{align*}
C_{t+\Delta} = C_{t} +  (E_{t+\Delta} - S_{t+\Delta}^{LND} - S_{t+\Delta}^{OCN}) \Delta, \qquad 
&S_{t+\Delta}^{OCN} =  -b_1  C_{1750} + b_1 C_t \exp(-c_1 T_t^d)   + \eta_{t+\Delta}^{O}, \\
& S_{t+\Delta}^{LND} = -b_2  C_{1750}  + b_2 C_t \exp(-c_2 T_t^d)  + \eta_{t+\Delta}^{L},
\end{align*}
where $b_1, b_2, c_1, c_2 \in \mathbb{R}$ are fixed unknown coefficients,
$C_{1750} = 591.3060$ GtC is the pre-industrial level of CO$_2$ concentrations,
$E_t = E_t^{FF} + E_t^{LUC}$, 
and disturbances $\eta_{t+\Delta}^{O}$ and $\eta_{t+\Delta}^{L}$ are mean-zero innovation terms. 
%
The forcing equation is discussed in Section \ref{sec:Feq1st} and we consider the
equations \eqref{eq:F}--\eqref{eq:F2} as given by
\[
F_{t+\Delta}^{CO2} = g_{F}(C_t) - g_{F}(C_{1750}) + \eta_{t+\Delta}^F, \qquad   \qquad
g_{F}(C_t) = f_1\log \left(C_t +f_2 C_t^2\right)  + f_3\sqrt{C_t},
\]
with mean-zero innovation $\eta_{t+\Delta}^F$ and $f_1, f_2, f_3 \in  \mathbb{R}$ are fixed unknown coefficients. 
%
The energy balance module equations are discussed in Section \ref{sec EBM} and are given by
\begin{align*}
T_{t+\Delta}^m &= \left(1-\frac{\gamma+\lambda}{H_m} \Delta \right)T^m_t + \frac{\gamma \Delta}{H_m} T^d_{t} + \frac{\Delta}{H_m}(F_t^{CO2} + F_t^{Ex})  + \eta_{t+\Delta}^m, \\
T^d_{t+\Delta} &= \frac{\gamma \Delta}{H_d}   T^m_t + \left( 1 - \frac{\gamma \Delta}{H_d} \right)T^d_t + \eta_{t+\Delta}^d, 
\end{align*}
where $\gamma, \lambda, H_m, H_d \in \mathbb{R}$ are fixed unknown coefficients,
$F_t^{Ex} = F_t^{Non-CO2} + F_t^{Nat}$ is forcing from non-CO$_2$ sources
(anthropogenic and natural), which are treated here as covariates,
and disturbances $ \eta_{t+\Delta}^{m}$ and $\eta_{t+\Delta}^{d}$ are mean-zero innovation terms. 


\subsection{Measurement equations}\label{sec:measurement}

We assume that we have a number of observed variables available that we can regard as, possibly noisy, measurements
for the state variables. We will denote most of these observed variables by the same name as their
state variable counterparts but we affix an asterisk to it.
Many different observed variables can be considered and the core selection typically
depend on data availability.
In our study, we consider the following set of seven observed variables with associated equations,
\begin{align*}
C_t^* &= \mu_C + C_t + \epsilon_t^C, \quad &\textnormal{(CO2 concentrations)} \\
S_t^{OCN,*} &= \mu_O + S_t^{OCN}  + \epsilon_t^{OCN}, \quad &\textnormal{(CO2 ocean sink)} \\
S_t^{LND,*} &= \mu_L + S_t^{LND}  + \epsilon_t^{LND}, \quad &\textnormal{(CO2 land sink)} \\
F_t^{CO2,*} &= \mu_F + F_t^{CO2} + \epsilon_t^F,  \quad &\textnormal{(Forcing from CO2)} \\
T_t^{m,*} &= \mu_m + T^m_t + \epsilon_t^m, \quad &\textnormal{(Surface temperature)} \\
T_t^{d,*} &=  \mu_d+  T^d_t + \epsilon_t^d,  \quad &\textnormal{(Deep ocean temperature $0$--$2000$m)} \\
O_t^* &= H_d \cdot \mu_d + H_d \cdot T^d_t + \epsilon_t^O,  \quad &\textnormal{(Ocean heat content $0$--$2000$m)} 
\end{align*}
where $H_d$ is the fixed unknown coefficient from the $T^d_{t}$ state equation in Section \ref{sec:state}, 
all $\epsilon^\cdot_t$ variables are mean-zero error processes and the unknown fixed constants
$\mu_\cdot \in \R$ are introduced to ensure that all state variables, i.e. the
variables without asterisks, are benchmarked to the same base period.
In our application below, we will choose this to be the pre-industrial period,
represented by the year $1750$.
This means that $\mu_m$, for instance, denotes the offset needed in the observations
$T^{m,*}$ to benchmark these to the pre-industrial period, i.e. $T^{m,*} - \mu_m$ is
the surface temperature anomaly relative to this period.
The last measurement equation for ocean heat content $O_t^*$ is obtained
from the relationship $$H_d \frac{dT^d_t}{dt} = \frac{dO_t}{dt},$$
\citep[e.g.][]{Schwartz2007,BHL2022} and the fact that the observation time series $T_t^{d,*}$ and $O^{*}_t$ are anomalies assumed to be benchmarked to the same historical period.

\subsection{State space representation}\label{sec:system}
Let $y_t = (C_t^*,S_t^{OCN,*},S_t^{LND,*},F_t^{CO2,*},T_t^{m,*},T_t^{d,*},O_t^*)'$ denote
the 
$7\times 1$
vector of observations at time $t$.
The resulting non-linear state space model for the climate system described in Section \ref{sec:model}
is
\[
y_t          \ = \ \mu + A x_t + \epsilon_t, \qquad  \qquad
x_{t+\Delta}   \ = \ B(x_t) + W_t +  R \eta_{t,\Delta},
\]
where the $7\times 1$ vector $\mu$ contains the intercepts in the measurement equations of Section \ref{sec:measurement}, the 
$7\times 6$ matrix $A$ enables the appropriate selection of the element in the
$6 \times 1$ state vector $x_t = (C_t, S_t^{OCN}, S_t^{LND}, F_t^{CO2}, T_t^{m}, T_t^{d})'$ for
each variable in $y_t$, the $7\times 1$ vector $\epsilon_t$ contains the measurement errors
listed in Section \ref{sec:measurement}, 
the $6\times 1$ non-linear vector function $B(\cdot)$ captures the
non-linear dynamic equations for the latent state variables in Section \ref{sec:state}, the 
$6\times 1$ vector $W_t$ consists of zeros and composite covariates
such as $\Delta(F_t^{CO2}+F_t^{Ex})/H_m$, and the 
$6\times 5$ matrix $R$ selects the appropriate element from the
$5\times 1$ innovation vector $\eta _{t,\Delta}=(\eta _t'^{O},\eta _t^{L},\eta _t^{F},\eta _t^{m},\eta _t^{d})'_{\Delta}$.
The state update equation comprises the dynamic processes of the physical quantities in Section \ref{sec:state}.
The measurement error sequence $\epsilon_t$ captures the transitory deviations
between observations and the corresponding underlying state processes.
We allow each element in $\epsilon_t$ to be a serially correlated sequence
modelled as a first-order stationary autoregressive process, that is
$$\epsilon_{t+\Delta} = \Phi \epsilon_{t} + \xi_{t,\Delta}, \qquad \xi_{t,\Delta} \stackrel{iid}{\sim} N(0,\Delta \cdot P),$$
where the autoregressive coefficient matrix $\Phi$ is diagonal, $N(0,V)$ refers to the multivariate normal distribution with a zero mean vector and
variance matrix $V$, and $iid$ means \emph{independent and identically distributed}.
To impose the stationarity of $\epsilon _t$, we assume that each diagonal element of $\Phi$ is smaller than unity in absolute value.
Given that variance matrix $P=Var(\xi _{t,\Delta})/\Delta$, stationarity implies that
$Var(\epsilon _t) = (I-\Phi^2)^{-1}P$.
A detailed discussion on the dynamic specification of $\epsilon _t$ is presented in Section \ref{sec GETS}.
We further have the innovation sequence
$$\eta _{t,\Delta}  \stackrel{iid}{\sim} N(0,\Delta \cdot Q),$$
with diagonal variance matrix $Q$,
in recognition that the physical equations in Section \ref{sec:state}
are only approximations to the true climate system.
It follows that $R \eta_{t,\Delta} = x_{t+\Delta}-B(x_t)-W_t$ can be regarded
as the error between the ``true'' climate system and our simplified non-linear version of it. 
More details on the entities $\{ A, \, B(\cdot), \, R, \, \Phi, \, P, \, Q\}$, and how they are
specified in the Stat-RCM, are provided in Appendix \ref{app:ssr}.
The fixed unknown constants and coefficients are collected in the parameter vector $\theta$
and are estimated
by the method of maximum likelihood, where
the extended Kalman filter is used to evaluate the likelihood function.
We refer to \citet[][Chapter 10]{durbin2012time} for further details on this approach to estimation.

\section{Monte Carlo simulation study}\label{sec:sims}

We perform a Monte Carlo simulation study designed to gauge the finite sample performance
of the maximum likelihood estimation method in a controlled setting. 
We use the historical CO$_2$ emissions and forcing data over the period $1959$--$2022$
as the covariates in the Stat-RCM model;
see Section \ref{sec:empirical} for a more in-depth discussion on these data. 
We then simulate the stochastic processes $\epsilon_t$ and $\eta_t$ from their specifications
provided in Section \ref{sec:system}, and
use these to generate simulated paths for the state vector $x_t$ and
the observation vector $y_t$ over the period $1959$--$2022$,
resulting in $n = 64$ observations.
The model parameters used for simulation are set equal to the estimates
we have obtained from our empirical study in Section \ref{sec:empirical}.
In particular, we have set $\mu_C = \mu_L = \mu_O = \mu_F = 0$ and $f_2 = f_3 = 0$.
The remaining parameters are estimated freely by numerically maximizing
the likelihood function obtained by the extended Kalman filter.
We repeat this simulation-estimation process $1000$ times to obtain $1000$
Monte Carlo estimates of the parameters of the model.

\begin{figure}[!t] 
\centering 
\includegraphics[width=0.95\columnwidth]{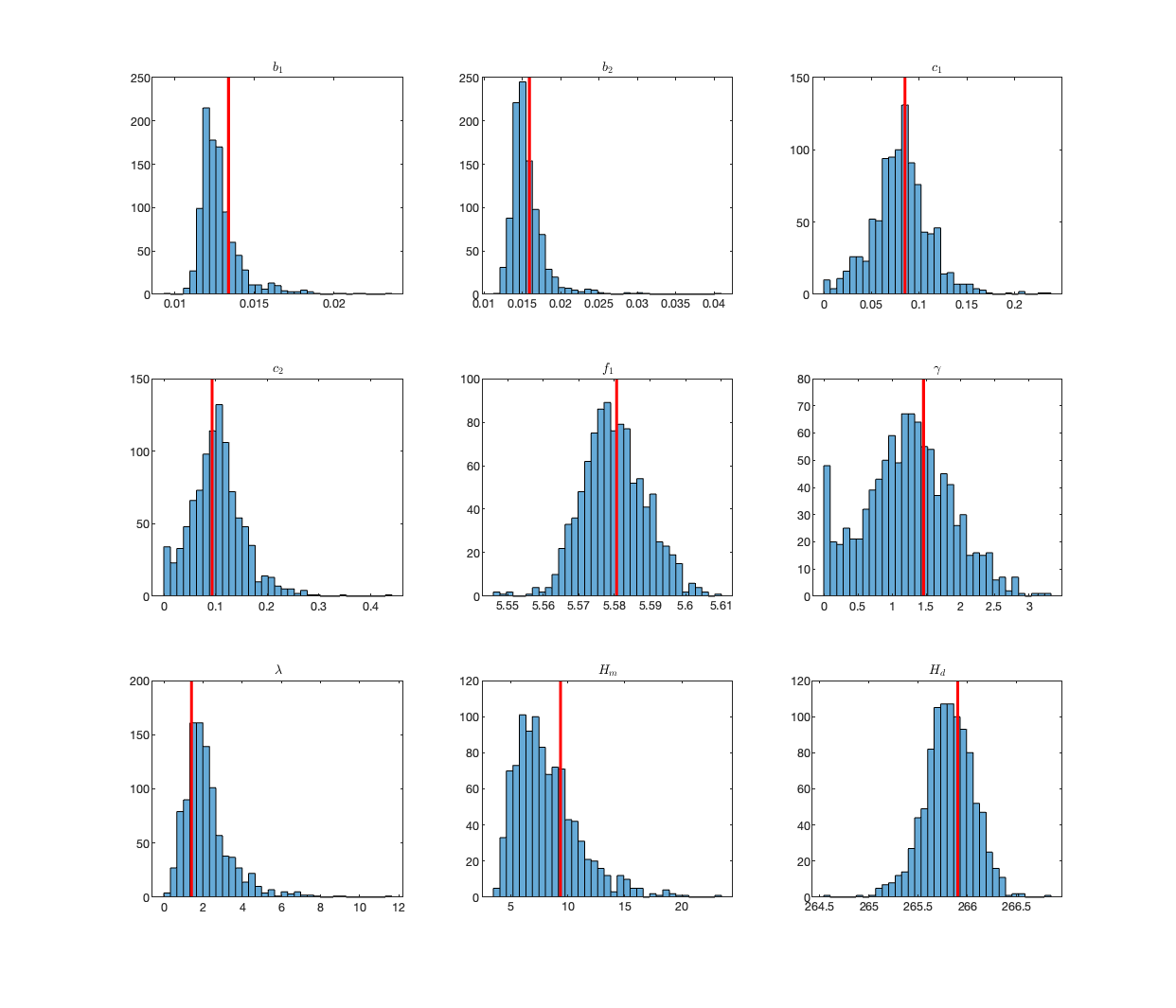}
\caption{\it Histogram of the parameter estimates of $\tilde \theta = (b_1,\, b_2,\, c_1,\, c_2,\, f_1,\, \gamma, \, \lambda, \, H_m, \, H_d)'$ obtained from $1000$ Monte Carlo replications, see Section \ref{sec:sims}. The vertical red line indicates the true parameter used as data generating value in the simulations.}
\label{fig:simRes}
\end{figure}

Table \ref{tab:simRes1959} summarizes the results of the simulation experiment,
where we focus on the estimates of the physical parameters in the model,
$\tilde \theta = (b_1,\, b_2, \, c_1, \, c_2, \, f_1,\, \gamma, \, \lambda, \, H_m, \, H_d)'$,
and the constant offsets $\mu_m$ and $\mu_d$.
Figure \ref{fig:simRes} presents histograms of the $1000$ estimates for each
of the parameters in $\widetilde \theta$, where a vertical red line indicates
the `true' data-generating value of the respective parameter.
We can conclude that the parameters of the carbon cycle module
$\{b_1,\, b_2,\, c_1,\, c_2\}$ and forcing equation $\{f_1\}$
are estimated with a high degree of precision.
The parameters from the energy balance module $\{\gamma,\, \lambda,\, H_m,\, H_d\}$,
as well as the constant offsets $\{\mu_m,\, \mu_d\}$, appear more difficult to estimate.
In particular, the estimates of $H_m$ and $H_d$ appear to show some small-sample bias.
Overall, the finite sample properties of the estimation procedure are adequate
in this Monte Carlo study design and this choice of parameter values. 


\begin{table}
\caption{\it Monte Carlo simulation results}
\begin{center}
\tiny
\begin{tabularx}{1.00\textwidth}{@{\extracolsep{\stretch{1}}}l|ccccccccccc@{}} 
\toprule
 & $b_1$ & $b_2$ & $c_1$ & $c_2$  & $f_1$ & $\gamma$ & $\lambda$ & $H_m$ & $H_d$ & $\mu_m$  & $\mu_d$    \\  
\midrule
True values: & $   0.01 $ & $   0.02 $ & $   0.09 $ & $   0.09 $ & $   5.58 $ & $   1.44 $ & $   1.44 $ & $   8.97 $ & $ 265.88 $ & $   0.28 $ & $ 0.05   $\\
\cmidrule{2-12}   
MC Avg & $   0.01 $ & $   0.02 $ & $   0.08 $ & $   0.10 $ & $   5.58 $ & $   1.24 $ & $   2.21 $ & $   8.23 $ & $ 265.81 $ & $   0.39 $ & $   0.11 $\\
(MC Std.)  & $(   0.00 )$ & $(   0.00 )$ & $(   0.03 )$ & $(   0.05 )$ & $(   0.01 )$ & $(   0.64 )$ & $(   1.25 )$ & $(   2.81 )$ & $(   0.25 )$ & $(   0.17 )$ & $(  15.78 )$\\
\bottomrule 
\end{tabularx}
\end{center}
{\footnotesize \it Simulation results for physical parameters
$\tilde \theta = (b_1,\, b_2,\, c_1,\, c_2,\, f_1,\, \gamma,\, \lambda, \, H_m, \, H_d)'$
as well as the constant offsets $\mu_m$ and $\mu_d$.
Number of Monte Carlo replications is $1000$ and number of observations is $n = 64$.
The first row reports the true parameters used for simulating the paths of the state
and measurement variables and the second and third row report the Monte Carlo average
and Monte Carlo standard deviation, respectively, calculated over the $1000$ replications.} 
\label{tab:simRes1959}
\end{table}

\section{The performance of Stat-RCM on the $1959$-$2022$ historical record}\label{sec:empirical}

We consider a historical yearly time series data set from $1959$ to $2022$.
The data for the carbon cycle
$\{C_t^*, \, S_t^{OCN,*}, \, S_t^{LND,*}, \, E_t^{FF}, \, E_t^{LUC}\}$
are from the Global Carbon Project \citep{GCB2022short}\footnote{\url{https://www.globalcarbonproject.org}.}. 
The forcing data $\{F_t^{CO2,*},\, F_t^{Non-CO2},\, F_t^{Nat}\}$ are the
current best estimates taken from the Sixth Assessment Report (AR6) of the
IPCC \citep{ipcc2021supp7}\footnote{\url{https://zenodo.org/record/5705391\#.ZFIx_y9ByX2}.}. 
The global mean surface temperature data ($T_t^{m,*}$) are from  HadCRUT5 by the Met Office Hadley Centre \citep{morice2020updated}\footnote{\url{https://www.metoffice.gov.uk/hadobs/hadcrut5/data/HadCRUT.5.0.2.0/download.html}.}. 
The deep ocean temperature data ($T_t^{d,*}$) and the ocean heat content data ($O_t^*$)
are from the Institute of Atmospheric Physics \citep[IAP;][]{Cheng2017}\footnote{\url{http://www.ocean.iap.ac.cn/}, all data resources 1-4 are last accessed on March 13, 2024.}. 

The AR6 forcing data of \cite{ipcc2021supp7} end in $2019$ while our remaining data run until $2022$.
For the forcing variable $F_t^{CO2}$, which is a model variable, it does not pose an issue
since the Kalman filter and smoother can easily handle missing data values.
For the covariates, including the forcing variables $F_t^{Non-CO2}$ and $F_t^{Nat}$, however, we need a complete data record. We therefore impute the $2020$--$2022$ values of $F_t^{Non-CO2}$ and $F_t^{Nat}$ using a linear trend, estimated from the last five years of the data, i.e. $2015$--$2019$. The results obtained from the Stat-RCM, reported below, are robust to alternative approaches to filling in these missing values. This is not surprising since the forcing from CO$_2$ is much larger in magnitude than the forcing from non-CO$_2$ greenhouse gases and natural forcing.


The carbon cycle data from the Global Carbon Project and the forcing data for the AR6 report are benchmarked to $1750$, meaning that we may set $\mu_C = \mu_L = \mu_O = \mu_F = 0$. We benchmark the surface temperature data, $T_t^{m,*}$, to the base period $1850$--$1900$ and the deep ocean data, $T_t^{d,*}$ and $O_t^*$, to $1940$.  To ensure that the state variables $T^m_t$ and $T_t^d$ are benchmarked to $1750$, we include the offset parameters $\mu_m$ and  $\mu_d$ in the relevant measurement equations and estimate these freely.
%

\subsection{The dynamic specification of the error structure}\label{sec GETS}

In Section \ref{sec:system}, we have modelled the stochastic error variables in
the $7 \times 1$ vector $\epsilon _t$ as autoregressive processes of order 1.
This type of process falls into the class of the autoregressive moving average (ARMA)
processes which is well-known in time series analysis \citep{BD1996}.
The ARMA model can have different maximum lag lengths for the autoregressive part, denoted by $p$, and
for the moving average part, denoted by $q$.
We first have considered the ARMA specification, with $p=q=1$, for each error term in $\epsilon _t$.
In addition, we have included a correlation coefficient, $\rho$, between the ARMA innovation
term $\xi _t$ (see also Section \ref{sec:system}) associated with the measurement
equations for $T^d_t$ and $O_t$. These two variables are closely related and therefore
this dependence is introduced, see also \cite{BHL2022}.
The initial estimates of the ARMA parameters have revealed
that parameters of the moving average part are
statistically insignificant at the $10\%$ level, while those of the autoregressive part are
significant. We therefore have opted for the autoregressive part only and
obtained the specification for $\xi _t$ as in Section \ref{sec:system},
also see Appendix \ref{app:ssr} for further details. 
The estimate of the dependence coefficient $\rho$ is
statistically significant and therefore it is kept in the specification for $\xi _t$.

\subsection{Model selection: the forcing equation}\label{sec:modelselection}

The Beer-Lambert law suggests that the functional relationship between atmospheric CO$_2$ concentration and radiative forcing from CO$_2$ is approximately logarithmic \citep[][]{LM2014}. Various functional forms have been suggested to improve upon this approximation \citep[e.g.][]{Hansen1988,Shi1992,Hansen1998}.
As discussed in Section \ref{sec:Feq1st}, there appears to be no consensus in the RCM literature on which of these functional forms is most adequate for describing the link between atmospheric concentrations and radiative forcings. Here, we will adopt statistical model selection methods to determine which functional form is most appropriate for the historical data under study.
To the best of our knowledge, statistical model selection methods have not been used to inform the construction of RCMs previously.
Given the statistical formulation of the Stat-RCM, model selection is relatively straightforward to carry out. 

In relation to the forcing equation \eqref{eq:F2},
that is $g_{F}(C_t) = f_1\log \left(C_t +f_2 C_t^2\right)  + f_3\sqrt{C_t}$,
we are interested in testing the hypotheses
\begin{align*}
H_0^{sqrt}: \ f_1 &= f_2 = 0, \\
H_0^{log}: \ f_2 &= f_3 = 0, \\
H_0^{log+sqrt}: \ f_2 &= 0, \\
H_0^{log2}: \ f_3 &= 0, \\
H_0^{Hansen98}: \ f_1 &= 5.04, \ f_2 = 0.00023507, \ f_3 = 0.
\end{align*}
The two hypotheses $H_0^{sqrt}$ and $H_0^{log}$ can be used to test whether the
forcing equation reduces to a square root function and a logarithmic function,
respectively. The hypothesis $H_0^{log}$ reduces the forcing equation to the one used in MAGICC6 \citep[][p. 1439]{MAGICC}. Under the hypothesis $H_0^{log+sqrt}$, the forcing equation is the sum of a square root term and a logarithmic term, which is the forcing equation used in the FaIR RCM \citep[][p. 3013]{FAIR}. The forcing equation implied by $H_0^{log2}$ is proposed in \cite{Hansen1998}. The hypothesis $H_0^{Hansen98}$ implies the same forcing equation, but where the parameters are fixed to the specific values used in \cite{Hansen1998}.\footnote{Table 1 of \cite{Hansen1998} actually specifies the forcings function $g_{F}(c_t) = 5.04 \log(c_t  + 0.0005 c_t^2)$, where $c_t$ is atmospheric concentrations given in ppm. Since $C_t = c_t \cdot 2.127$GtC/ppm, this translates into the parameters $f_1 = 5.04$ and $f_2 = 0.0005/2.127 =0.00023507$ in our framework. When estimating the parameters from this forcing equation, i.e. when working under $H_0^{log2}$, we find  $\hat f_1 =  5.58$ and $\hat f_2 =  0.00000105$, indicating that the forcing data used in this paper results in somewhat different parameter estimates than what was used in \cite{Hansen1998}.} See also \citet[][Table 6.2., p. 358]{Ramaswami2001} for various suggestions for expressions of the forcing equation, including some of those studied here.

Since the models implied by the hypotheses listed above are nested in
the general forcing equation \eqref{eq:F2}, where $f_1$, $f_2$ and $f_3$ are unrestricted coefficients,
we can perform the various hypothesis tests using a standard likelihood ratio test.
Table \ref{tab:FlikRatio} contains the maximized log-likelihood values
for the models with different forcing specifications, together with
the Bayesian Information Criteria \citep[BIC;][]{BIC1978},
the numbers of estimated parameters,
and the $p$-values for the likelihood ratio tests.
We find that the model specification under $H_0^{log}: \ f_2  = f_3= 0$ above
has the lowest BIC value and hence has most support from the data.
The two hypotheses $H_0^{sqrt}$ and  $H_0^{Hansen98}$  can formally be
rejected at a $1\%$ significance level and $H_0^{log2}$ barely at a $5\%$ level.
The rejection of $H_0^{sqrt}$ indicates that a log-term is preferred
to a square root term when modelling the relationship between
atmospheric CO$_2$ concentrations and radiative forcing,
for this updated AR6 forcing data set.
Finally, the rejection of $H_0^{Hansen98}$ indicates that the parameter values
used in \cite{Hansen1998} do not fit this data set.

The reported results below are based on the model specification
under $H_0^{log}$ as it produces the lowest BIC value,
and is not rejected by the likelihood ratio test.
The model specifications with a comparable fit as the $H_0^{log}$ model, that is the model implied by $H_0^{log+sqrt}$ and $H_0^{log2}$ and the unrestricted model,
yield similar results as those from the model with $H_0^{log}$.

\begin{table}
\caption{\it Likelihood ratio tests for the forcing equation}
\begin{center}
\scriptsize
\begin{tabular}{l|ccccccc@{}} 
\toprule
                                                   & Unrestricted       & $H_0^{sqrt+log}$  & $H_0^{log2}$       & $H_0^{sqrt}$         & $H_0^{log}$       & $H_0^{Hansen98}$  \\  
\midrule
Maximized log-likelihood:          & 925.79         &    924.78         &       923.77      &    917.17         &   923.76      &    911.21   \\
Bayesian Information Criterion:    & -1714.34       & -1716.48          &  -1714.45         &  -1705.41         & -1718.60      & -1697.67\\
Number of parameters:              & 33             & 32                &  32               & 31                & 31            & 30 \\
$p$-value (Likelihood Ratio test): &   -            &   0.1552          &  0.0444           &  0.00018          &    0.1313     &   0     \\
\bottomrule 
\end{tabular}
\end{center}
{\footnotesize \it  The Likelihood Ratio tests are against the ``unrestricted'' model with coefficients $f_1$, $f_2$ and $f_3$ estimated freely.} 
\label{tab:FlikRatio}
\end{table}



\subsection{Estimation results and assessment of model fit}\label{sec estimation results}

Given the findings from the model selection exercise above, we work under the hypothesis
$H_0^{log}$, i.e. we set $f_2 = f_3 = 0$ and estimate the parameter $f_1$ in the forcing equations
\eqref{eq:F}--\eqref{eq:F2} together with the other parameters.
The estimation results from this model applied to the 1959--2022 data are presented in Table \ref{tab:M7pI}.
For purposes of brevity, we only present the estimates from the ``physical'' parameters in
the model, i.e. the parameters governing the main workings of the carbon cycle module,
$\{b_1,\, b_2,\, c_1,\, c_2\}$,
the forcing module, $f_1$,
the energy balance module, $\{\gamma,\, \lambda,\,  H_m,\, H_d\}$,
as well as the constant offsets from the measurement equations $\{\mu_m, \, \mu_d\}$.
Estimates of the remaining parameters are presented in Appendix \ref{app additional sims}.



The signs and magnitudes of the estimated carbon cycle feedback parameters $\{b_1,\,b_2\}$
are as anticipated and are found in previous studies \citep[e.g.][]{BHK2023b}.
The climate feedback parameters $\{c_1,\, c_2\}$
are both estimated as $\widehat c_1, \widehat c_2 \approx 0.09$, implying that
a temperature rise of $1^\circ C$, above pre-industrial levels,
would correspond to a climate feedback factor of
$\exp(-  T_t^m \cdot \widehat c_i) = \exp(-1\cdot 0.09) \approx 0.9139$, for $i=1,2$.
This is roughly a $9\%$ weakening of the sinks compared to pre-industrial
levels (notice that $T_{1750}^m = 0$).
A similar magnitude of the climate feedback estimate obtained from historical data has been reported previously \citep[][]{ZhangX2021} but it is lower than values
found using large-scale climate models \citep[][]{Friedlingstein2015}
The reason for the low estimate of the climate feedback effect
in the Stat-RCM is probably due to the input data used,
i.e. the historical data $1959$--$2022$, where climate feedback effects
are still rather weak.
To obtain more precise estimates of the climate feedback effect would
require data where these feedbacks are more manifest.
This could for instance be achieved by fitting the Stat-RCM to output from
large-scale climate models run over longer periods than the historical period,
similar to what is done for other RCM emulators \citep[][]{RCMIP2020}.
This avenue is left for future research.

In Table \ref{tab:M7pI}, we present the results for the forcing module and we find that
the estimated parameter $\widehat f_1$ is highly significant and has the expected sign.
The parameter estimates from the energy balance module align well with previous studies
\citep[][]{Cummins2020,Pretis2020,BHL2022}.
We notice that an estimate of the equilibrium climate sensitivity (ECS) can be derived
from our estimate of the climate feedback parameter $\lambda$.
In particular, the IPCC Sixth Assessment Report \citep[AR6;][]{ipcc2021c7}
suggests the relationship $ECS = F_{2\times CO2} / \lambda$, where $F_{2\times CO2}$ is
the radiative forcing in response to a doubling of CO$_2$ concentrations in the atmosphere.
Using the best estimate of
$\widehat F_{2\times CO2} \approx 3.93$ ($\pm 0.47, \ 5\%$--$95\%$ CI) Wm$^{-2}$ from AR6 (updated),
we find that $\widehat{ECS} = F_{2\times CO2} / \widehat \lambda = 3.93 / 1.42 = 2.78^\circ C$.
This estimate aligns well with the ``likely'' range of
$2.5^\circ$--$4.0^\circ C$ obtained from instrumental records and reported in AR6
\citep[][]{ipcc2021c7}.
Under certain assumptions, 
we can also obtain the
(asymptotic) variance of $\widehat{ECS}$
which results in a standard error of $1.03$.\footnote{The
standard error of $\widehat{ECS}$ can be obtained as follows.
Define $\widehat{ECS} = g(\widehat F_{2\times CO2},\widehat \lambda) = \widehat F_{2\times CO2} / \widehat \lambda$,
where $g(x,y)$ is the function $g(x,y) = x/y$.
The Jabobian of $g$ is $\partial g = (1/ \widehat \lambda,-\widehat F_{2\times CO2} / \widehat \lambda^2)'$.
Let $\Sigma$ be the $2\times 2$ diagonal matrix with $Var(\widehat F_{2\times CO2})$ and $Var(\widehat \lambda)$ on the diagonal.
An estimate of $Var(\widehat \lambda)$ can be obtained from the maximum likelihood method discussed in Section \ref{sec:system}. 
Assuming Gaussian distributions and a $90\%$ confidence interval of $( 3.93\pm 0.47)$
as reported in \cite{ipcc2021c7}, we obtain the estimate $\widehat{Var}(\widehat F_{2\times CO2}) = (0.47/1.6449)^2$.
The variance of $\widehat{ECS}$ can then be approximated via the `delta method' as $\widehat{Var}(\widehat{ECS}) = (\partial g)' \Sigma \partial g$.}
Hence, the Stat-RCM estimate and associated standard error of equilibrium climate sensitivity is
$\widehat{ECS} = 2.78$ with standard error of $1.03$.

\begin{figure}[!t] 
\centering 
\includegraphics[width=0.95\columnwidth]{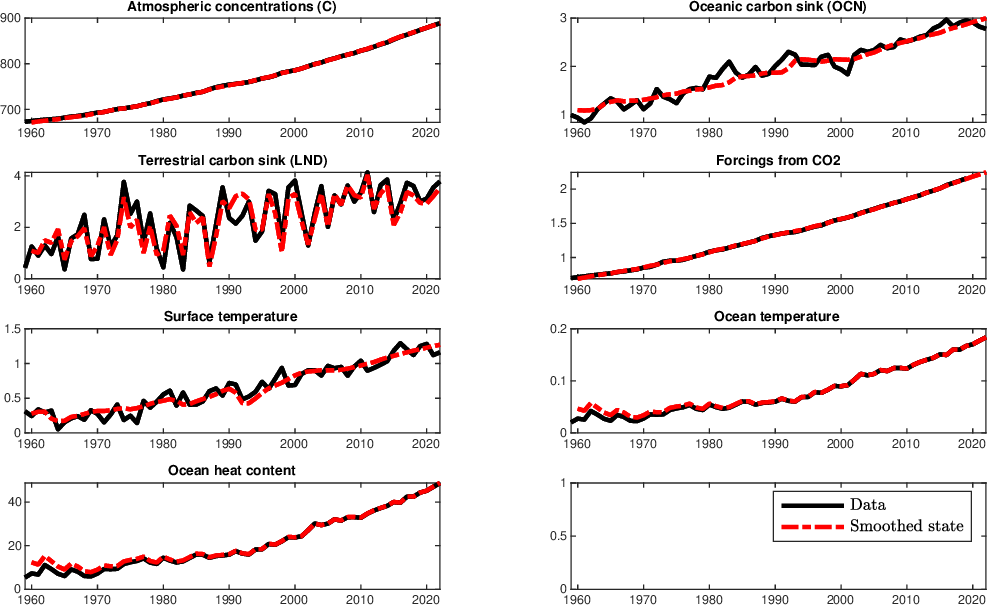}
\caption{\it Smoothed states from the Stat-RCM (red dashed lines), estimated on the historical data record $1959$--$2022$ (black lines), as output from the extended Kalman smoother. For surface temperature, ocean temperature, and ocean heat content, the estimated constant offsets ($\mu_m, \mu_d, H_d \cdot \mu_d$) have been added to the smoothed states to make them comparable to the data.}
\label{fig:M7states}
\end{figure}
\begin{figure}[!t] 
\centering 
\includegraphics[width=0.95\columnwidth]{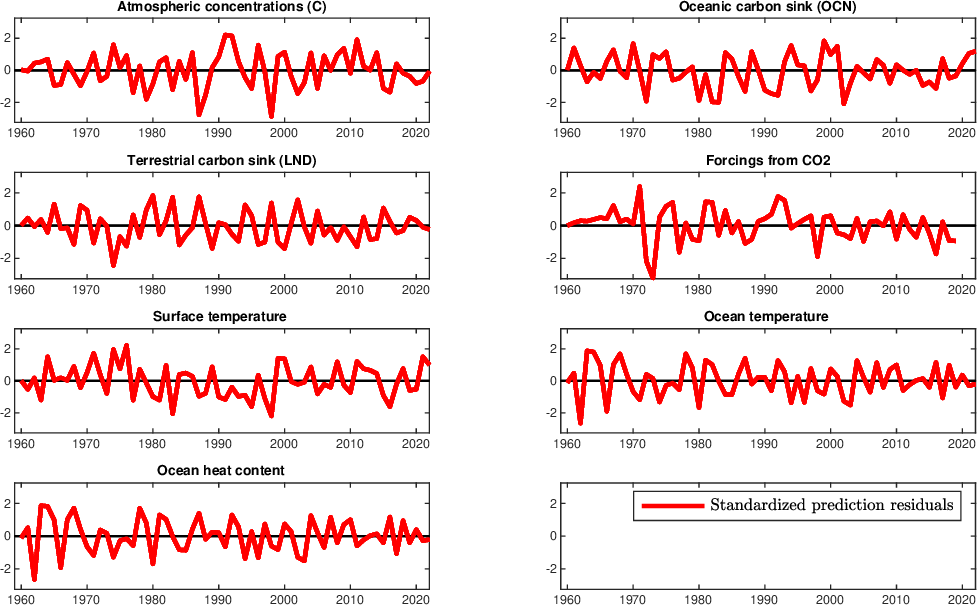}
\caption{\it Standardized one-step ahead prediction residuals from the Stat-RCM estimated on the historical data record $1959$--$2022$.}
\label{fig:M7res}
\end{figure}

For a correctly specified model, the standardized one-step ahead prediction residuals will be a sequence of iid $N(0,1)$ variables \citep{durbin2012time}. Diagnostics on these standardized prediction residuals are presented in Table \ref{tab:M7d}. There appears to be some autocorrelation left in the prediction residuals at the longer lags for the land sink.
The predictions residuals from the forcing equation display some indications of non-Gaussianity and heteroskedasticity due to outliers in the early $1970$s. Overall, however, the diagnostics are very reasonable and indicate that the model is able to fit the data well. The raw data, along with smoothed states as output from the extended Kalman filter, are presented in Figure \ref{fig:M7states}.
The standardized one-step ahead prediction residuals used for the residual diagnostics in Table \ref{tab:M7d} are displayed in Figure \ref{fig:M7res}.

\begin{table}
\caption{\it Stat-RCM parameter estimates}
\begin{center}
\tiny
\begin{tabularx}{1.00\textwidth}{@{\extracolsep{\stretch{1}}}l|ccccccccccc@{}} 
\toprule
 &   $b_1$ & $b_2$ & $\delta_1$ & $\delta_2$  & $f_1$  & $\gamma$ & $\lambda$ & $C_m$ & $C_d$   & $\mu_m$   & $\mu_d$    \\  
\midrule
Estimate:  & $   0.01 $ & $   0.02 $ & $   0.08 $ & $   0.09 $ & $   5.58 $ & $   1.46 $ & $   1.42 $ & $   9.37 $ & $ 265.90 $ & $   0.30 $ & $   0.20 $\\
Std. Err.:  & $(   0.00 )$ & $(   0.00 )$ & $(   0.02 )$ & $(   0.04 )$ & $(   0.01 )$ & $(   0.58 )$ & $(   0.51 )$ & $(   2.44 )$ & $(   0.41 )$ & $(   0.41 )$ & $(   0.36 )$\\
$t$-stat:  & $  11.25 $ & $   7.83 $ & $   5.16 $ & $   2.53 $ & $ 580.42 $ & $   2.51 $ & $   2.75 $ & $   3.84 $ & $ 656.40 $ & $   0.74 $ & $   0.55 $\\
\bottomrule 
\end{tabularx}
\end{center}
{\footnotesize \it Parameter estimates by the maximum likelihood method applied to the $1959$--$2022$ data.} 
\label{tab:M7pI}
\end{table}

\begin{table}
\caption{\it Residual diagnostics}
\begin{center}
\scriptsize
\begin{tabular}{c|ccccccccccc@{}} 
\toprule
 $y$   & Mean & Std. & Skew & Kurt & SC & JB & DW & LB$(1)$ & LB$(5)$ & LB$(10)$ & ARCH \\  
\midrule
C   & $  -0.01 $ & $   1.08 $ & $  -0.28 $ & $   3.02 $ & $   0.03 $ & $   0.81 $ & $   1.93 $ & $   0.08 $ & $   1.72 $ & $  12.17 $ & $   0.01 $\\
OCN   & $  -0.06 $ & $   1.00 $ & $  -0.18 $ & $   2.35 $ & $   0.14 $ & $   1.45 $ & $   1.71 $ & $   1.07 $ & $   6.64 $ & $   8.51 $ & $   0.18 $\\
LND  & $  -0.07 $ & $   0.92 $ & $   0.17 $ & $   2.65 $ & $  -0.09 $ & $   0.62 $ & $   2.18 $ & $   0.64 $ & $  17.10 $ & $  47.91 $ & $   1.63 $\\
Forc  & $   0.02 $ & $   1.01 $ & $  -0.55 $ & $   3.94 $ & $   0.05 $ & $   5.20 $ & $   1.89 $ & $   0.13 $ & $   2.06 $ & $   6.51 $ & $   5.29 $\\
Temp   & $  -0.01 $ & $   1.00 $ & $   0.06 $ & $   2.40 $ & $  -0.01 $ & $   0.98 $ & $   2.01 $ & $   0.01 $ & $   7.19 $ & $  15.48 $ & $   0.01 $\\
O-Temp   & $   0.04 $ & $   0.99 $ & $  -0.25 $ & $   2.62 $ & $  -0.18 $ & $   1.05 $ & $   2.37 $ & $   2.27 $ & $  10.30 $ & $  17.38 $ & $   2.88 $\\
OHC   & $   0.04 $ & $   1.00 $ & $  -0.25 $ & $   2.63 $ & $  -0.18 $ & $   1.03 $ & $   2.37 $ & $   2.29 $ & $  10.39 $ & $  17.36 $ & $   2.86 $\\
\bottomrule 
\end{tabular}
\end{center}
{\footnotesize \it Diagnostics of standardized one-step ahead prediction residuals as output from the extended Kalman filter. SC is the estimate of the serial correlation coefficient $\phi$ in the regression $y_t = \phi y_{t-1} + \epsilon$; $JB$ is the Jarque-Bera test statistic \citep[][]{JarqueBera1987}: the null hypothesis of Gaussianity is rejected if $JB$ is larger than the $95\%$ critical value of $5.99$. $DW$ is the Durbin-Watson test statistic \citep[][]{Durbin1971}: $DW<2$ indicates positive serial correlation, $DW>2$ negative serial correlation; $DW=2$ indicates no serial correlation.  $LB(p)$ is the Ljung-Box $Q$ test statistic \citep[][]{LB1978} for autocorrelation, calculated using $p$ lags. The $95\%$ critical value of the LB test are $3.84, 11.07, 18.31$ for $p = 1,5,10$, respectively. ARCH is the test statistic of the \cite{Engle1988} test for heteroskadasticity. The $95\%$ critical value of the ARCH test is $3.8415$.} 
\label{tab:M7d}
\end{table}

\subsection{State estimation}

The Kalman filter methodology allows for the efficient estimation of the latent variables in the $7\times 1$ state vector $x_t$.
In particular, the state estimate obtained from the Kalman smoother is an estimate of
$E[x_t | y_{1959:2022}]$ for $t = 1959, 1960, \ldots, 2022$, where $x_t$ denotes the state at time $t$
and $y_{1959:2022}$ denotes the full historical data set.
These smoothed states are presented in Figure \ref{fig:M7states}.

The estimated state for the surface temperature $T_t^m$, i.e.
$\widehat T_t^m =  \widehat E[T_t^m | y_{1959:2022}]$ for $t = 1959, 1960, \ldots, 2022$,
can be seen as an estimate of the underlying long-term temperature anomaly, i.e.
an estimate where transitory effects, such as measurement errors, captured in our
model by the error process $\epsilon_t$, have been filtered out.
Thus, our estimate of the surface temperature state can serve as a broad indicator of the overall state of the climate.
In this way, it can be used to assess how close we are to breaching international temperature agreements such as
the 2015 Paris Agreement of keeping ``global temperatures well below $2^\circ$C above pre-industrial times
while pursuing means to limit the increase to $1.5^\circ$C'' \citep{FCCC2015}.
Assessment and detection of whether temperature targets have been breached is difficult,
due to the inherent noisiness and variability of temperature measurements \citep[][]{Betts2023}.
In AR6, the IPCC itself proposed a way of detection such as breach:
\emph{A breach is deemed detected if the average temperature over a $20$-year horizon exceeds the target} \cite[e.g.][]{AR6synthesis}. 
This means that there will be a long lag, on the order of a decade, before a breach of a temperature target
will be detected. Such slow detection times run the risk of delaying action that may be relevant in case of
a breach. For this reason, several methods have recently been suggested, which are able to monitor the
underlying global temperature level closer and offer faster detection times of possible breaches.
For instance, \cite{Betts2023} suggests to blend the past $10$ years of historical temperature data
with projections of the next $10$ years of temperature data, obtained from a climate model.
While this method is able to estimate a ``$20$-year average'' of the current temperature level without delay,
it has the downside that the credibility of the estimate depends on the projected temperature,
which in turn depends on the climate model used to construct the projections.
Alternative methods for keeping track of current underlying global temperatures include Copernicus'
approach of calculating a $30$-year linear trend into the future,\footnote{\url{https://cds.climate.copernicus.eu/apps/c3s/app-c3s-global-temperature-trend-monitor?month:float=10&year:float=2023}}
NASA's approach of using lowess smoothing,\footnote{\url{https://climate.nasa.gov/vital-signs/global-temperature/}}
and the ``Real-time Global Warming Index'' of \cite{Haustein:2017aa}, which estimates the amount of warming due to
anthropogenic causes by regressing temperature data on forcing time series. 
 
In contrast, the estimated temperature state from the Stat-RCM can be seen as a purely statistical and data-driven alternative to assess the current level of underlying warming above pre-industrial levels. This estimate is only based on historical data and it is compatible with the physical components of the Stat-RCM, i.e. the carbon cycle module, the forcing module, and the energy balance module. Based on the data used in this paper, the current best-estimate of the underlying warming above the $1850$--$1900$ baseline (the baseline usually used for defining temperature targets), obtained from the Stat-RCM, is $\widehat T_{2022}^m +\widehat \mu_m =\widehat E[T_{2022}^m | y_{1959:2022}] +\widehat \mu_m = 1.27^\circ$C. This number is substantially above the $2022$ temperature measurement, $T_{2022}^{m,*} = 1.16$, see Figure \ref{fig:M7states}.
This is mostly due to the $2022$ La Ni\~na conditions which have been filtered out
of the estimate $\widehat T_{2022}^m =\widehat E[T_{2022}^m | y_{1959:2022}]$, leading to a more precise representation of the underlying warming trend, i.e. a trend free from transitory deviations such as ENSO.
 


\section{Scenario projections using Stat-RCM}\label{sec:projections}

We compute scenario projections of the future climate system from the estimated Stat-RCM, conditional on
paths for the covariate (external) variables CO$_2$ emissions, forcing from non-CO2 greenhouse gases, and forcing
from natural sources.  We  focus particularly on the probabilistic projections.
Section \ref{sec:uncertainty} lays out the projection methodology, which is based on simulations
from the estimated Stat-RCM. Section \ref{sec:validation} contains a validation exercise,
where we use historical covariates from the period $1959$--$2022$ and investigate whether
the Stat-RCM is able to produce ``projections'' over $1959$--$2022$ that agree with the historical data record.
Section \ref{sec:netzero} presents scenario projections for $2023$--$2100$ in a setting where CO$_2$ emissions
are rapidly declining, a necessary condition for halting global warming and achieving the goals as set out
by international agreements, such as the Paris Agreement \citep[e.g.][]{SONBT2016,Luderer2018,TG2018}. 

\subsection{Epistemic and aleatoric uncertainty in climate projections} \label{sec:uncertainty}

To gauge the degree of uncertainty introduced by the separate stochastic parts of the Stat-RCM,
we consider the following four different approaches of using the model to project the climate system forward, conditional on  paths of all covariate variables.
\begin{enumerate}
\item\label{it:setup1}
    Deterministic run: Deterministic projection where the parameters of the Stat-RCM, $\theta$,
    are set to their maximum likelihood estimates as reported in Section \ref{sec:empirical}, and based
    on the $1959$--$2022$ data; we have $\theta = \hat \theta$ and all error processes are set to zero, $\epsilon_t = \eta_t = 0$ for all $t$.
\item \label{it:setup2}
    Uncertainty coming from unknown parameters ($\theta$): Projection where the physical parameters,
    $\tilde \theta = (b_1,b_2,c_1,c_2,f_1,\gamma, \lambda, H_m,H_d)'$, are randomly sampled from the asymptotic distribution, that is
    $\tilde \theta \sim N(\hat \theta, \widehat \Sigma)$, where $\hat \theta$ and $\widehat \Sigma$ are the
    maximum likelihood estimates of the physical parameters and their variance-covariance matrix, based on the
    $1959$--$2022$ data. The remaining parameters are set equal to their maximum likelihood estimates.
    The error processes are set to zero, i.e.  $\epsilon_t= \eta_t = 0$ for all $t$.
\item \label{it:setup3}
    Uncertainty coming from unknown parameters and state innovations ($\theta, \eta$):
    Similar to the setup in \ref{it:setup2}, but now adding the stochastic variation due to
    the innovation terms in the state equations $\eta$,
    and setting the measurement error processes to zero i.e.  $\epsilon_t = 0$ for all $t$.
\item \label{it:setup4}
    Uncertainty coming from unknown parameters, state innovations, and measurement errors ($\theta, \eta, \epsilon$):
    Similar to the setup in \ref{it:setup3}, but now adding the stochastic variation due to the measurement errors $\epsilon_t$.
\end{enumerate}

In the first setup, the projection is deterministic, conditional on estimated parameters and
input paths of the covariates. The next three setups contain stochastic variation
to increasing degrees: First, by considering only randomness arising from the
uncertainty in parameters ($\theta$), then also including uncertainty coming from the
evolution of the state equations ($\theta$ and $\eta$),
and finally also including the uncertainty coming from other processes,
such as measurement errors and ENSO effects ($\theta$, $\eta$ and $\epsilon$).
We refer to the uncertainty about parameters as ``epistemic'' uncertainty,
in the sense that this uncertainty reflects our ignorance of the
data generating process behind the climate system, while uncertainty stemming
from the error processes $\eta$ and $\epsilon$ may be referred to as ``aleatoric'',
in the sense that this uncertainty captures the internal random variability of
the climate system itself.
To construct projections and confidence bands from the model under the various
uncertainty specifications, we simulate $10^5$ different instances of the
variables in the model using given paths for the covariates and plot the
pointwise quantiles of $2.5\%$, $50\%$ and $97.5\%$.

\subsection{Validation exercise}\label{sec:validation}

\begin{figure}[!t] 
\centering 
\includegraphics[width=0.95\columnwidth]{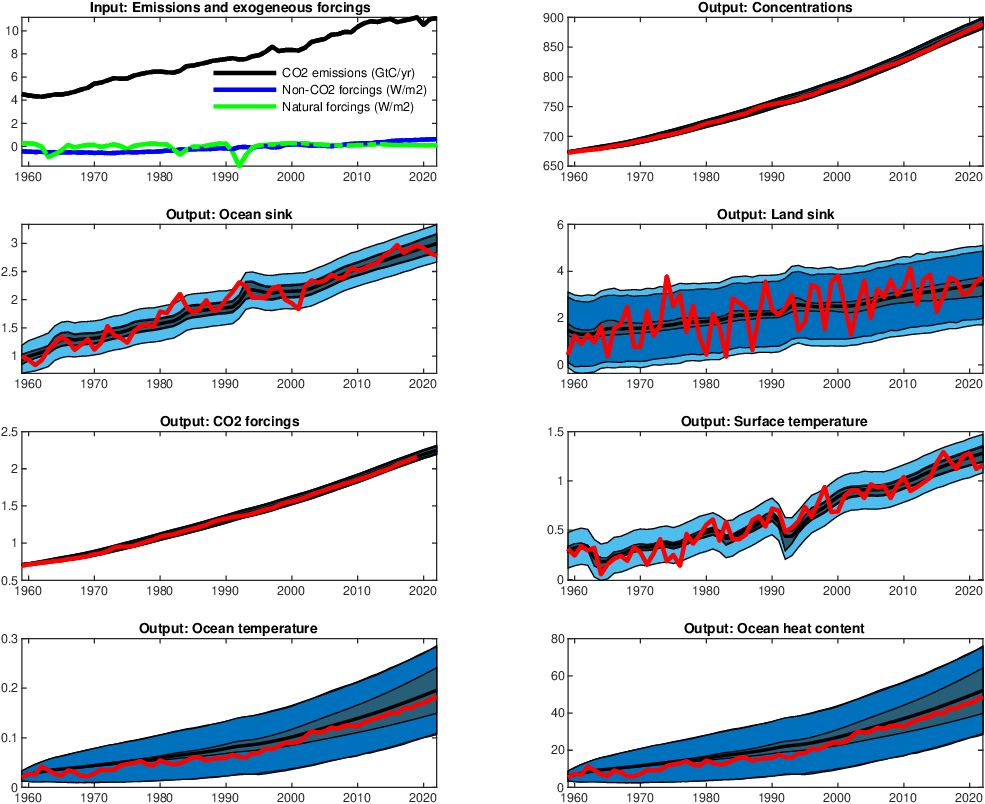}
\caption{\it Validation exercise using covariate data and estimated parameters from the historical data
$1959$--$2022$. Top left panel: Covariate input for the Stat-RCM. Remaining panels: Output of the Stat-RCM,
including data (red lines) for comparison. Black line: Path of deterministic run (Setup in \ref{it:setup1}). Dark grey shading: $95\%$ pointwise quantile bands for simulations with parameter uncertainty ($\theta$; Setup in \ref{it:setup2}). Dark blue shading:  $95\%$ pointwise quantile bands for simulations with uncertainty coming from parameters and innovations in the state equation ($\theta, \eta$; Setup in \ref{it:setup3}). Light blue shading: $95\%$ pointwise quantile bands for simulations with uncertainty coming from parameters, innovations in the state equation, and transitory error terms ($\theta, \eta, \epsilon$; Setup in \ref{it:setup4}). } 
\label{fig:M7valid_1}
\end{figure}

We adopt the model specification and its parameter estimates as reported in Section \ref{sec:empirical}
for the period $1959$--$2022$, to perform a
validation exercise. We take the covariates
CO2 emissions ($E_t$), forcing from non-CO2 sources ($F_t^{Non-CO2}$),
and forcing from natural sources ($F_t^{Nat}$) as inputs and keep them
fixed to their historical values.
The projection results are presented in Figure \ref{fig:M7valid_1}.
The top left panel shows the covariates (input variables) and the remaining panels show
the quantiles of the simulated model variables, along with the original observations
(red) used for the estimation of the model superimposed.
Although the original observations are used for the parameter estimates,
we stress that they are not otherwise used as input for this validation exercise.
From Figure \ref{fig:M7valid_1}, we find that the model is well-validated over
the estimation sample $1959$--$2022$, in the sense that it can produce simulations
of the climate system with a high degree of similarity to the observed historical data.

\subsection{Projections to $2100$ in a strong mitigation scenario}\label{sec:netzero}

\begin{figure}[!t] 
\centering 
\includegraphics[width=0.95\columnwidth]{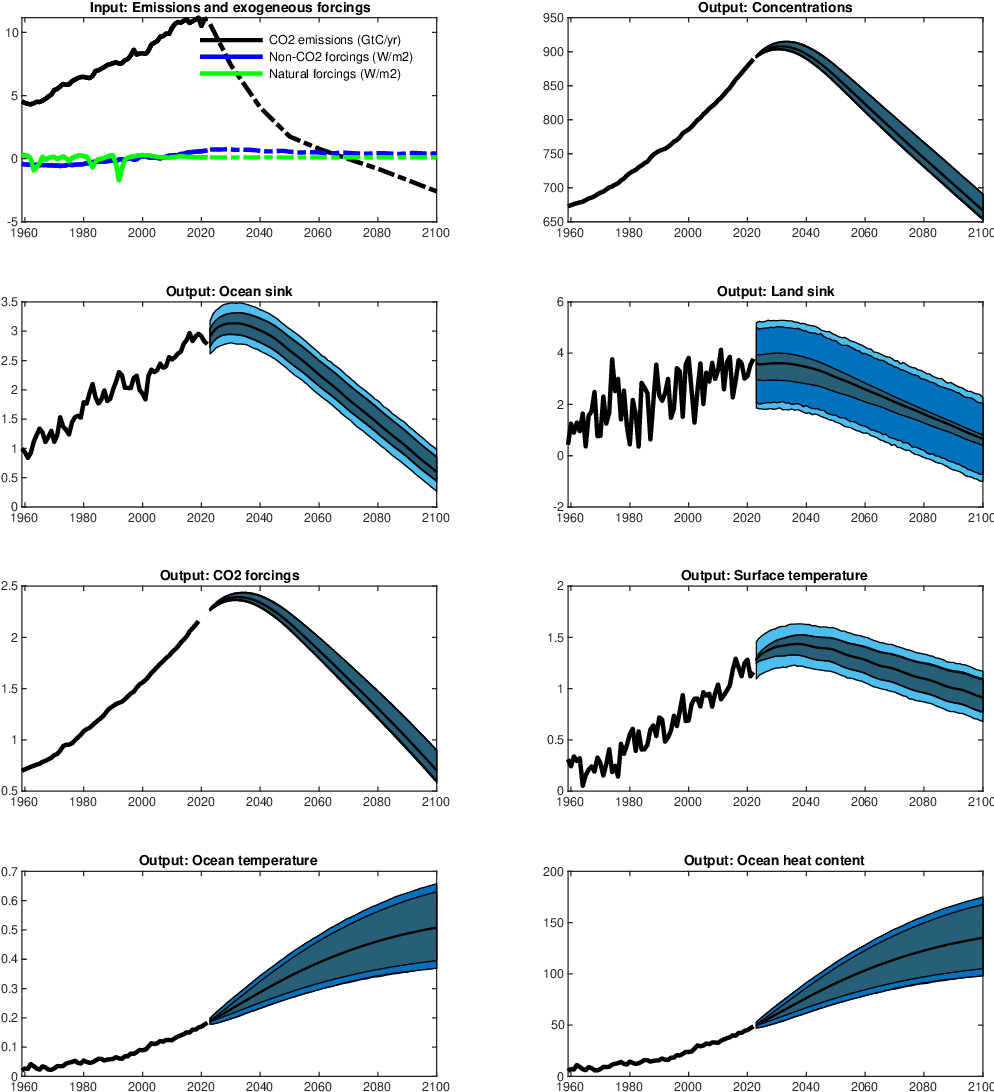}
\caption{\it Projections as based on covariates emissions and forcing data from SSP119, $2023$--$2100$, and Stat-RCM parameters estimated using historical data $1959$--$2022$. Top left panel: Covariates input to the Stat-RCM. Remaining panels: Output of the Stat-RCM, including data for comparison. Black line: Path of deterministic run (Setup in \ref{it:setup1}). Dark grey shading: $95\%$ pointwise quantile bands for simulations with parameter uncertainty ($\theta$; Setup in \ref{it:setup2}). Dark blue shading:  $95\%$ pointwise quantile bands for simulations with uncertainty coming from parameters and innovations in the state equation ($\theta, \eta$; Setup in \ref{it:setup3}). Light blue shading: $95\%$ pointwise quantile bands for simulations with uncertainty coming from parameters, innovations in the state equation, and transitory error terms ($\theta, \eta, \epsilon$; Setup in \ref{it:setup4}). } 
\label{fig:M7sims_scen_wUnc}
\end{figure}

Next, we use the Stat-RCM to project the variables until the year $2100$,
conditional on given future paths for the covariates of CO$_2$ emissions,
forcing from non-CO2 greenhouse gases, and natural forcing.
The path for CO2 emissions is taken from  the SSP119 scenario \citep[][]{SSP2020},
and forcings from non-CO2 greenhouse gases consistent with SSP119 are obtained
from the MAGICC RCM.\footnote{The SSP119 scenario can be run in MAGICC in a web
browser via the link \url{https://live.magicc.org/scenarios/bced417f-0f7f-4bb7-8359-792a0a8b0368/overview}.
Here, the forcing from non-CO2 greenhouse gases can also be downloaded.
Last accessed May 12, 2023.} The path for future natural forcing is set constant
equal to the last in-sample value.  The values for the future paths of the covariates
are shown in the top-left panel of Figure \ref{fig:M7sims_scen_wUnc},
where a solid line denotes historical data, from which the parameter estimates for Stat-RCM are obtained,
and dashed lines denote the future scenario used in the projection of the model.
We notice that the SSP119 scenario is very ambitious in the sense that it implies
rapid reductions in greenhouse gas emissions in the short term,
and substantial carbon dioxide removal from the atmosphere (negative emissions)
in the long term.

\begin{figure}[!t] 
\centering 
\includegraphics[width=0.95\columnwidth]{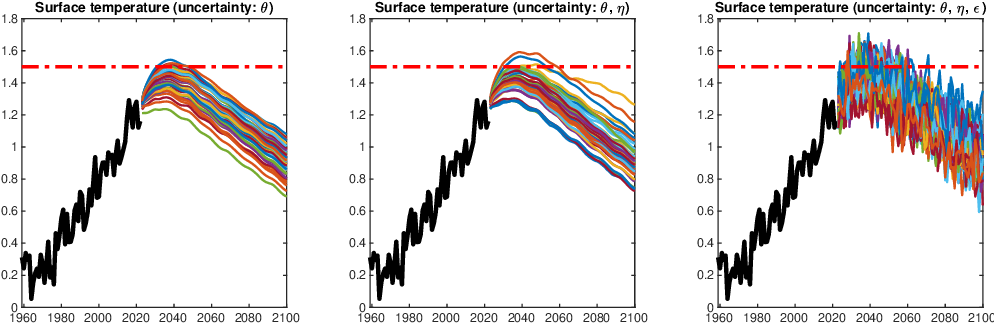}
\caption{\it $50$ simulated trajectories  of the surface temperature anomaly $T^m$, benchmarked to the period $1850$--$1900$. Black line: Historical data (1959--2022). Left panel: Simulations with parameter uncertainty ($\theta$; Setup in \ref{it:setup2}). Middle panel:  Simulations with uncertainty coming from parameters and innovations in the state equation ($\theta, \eta$; Setup in \ref{it:setup3}). Right panel:  Simulations with uncertainty coming from parameters, innovations in the state equation, and transitory error terms ($\theta, \eta, \epsilon$; Setup in \ref{it:setup4}). Red dashed line denotes $1.5^\circ$C. } 
\label{fig:M7sims_TAS_wUnc}
\end{figure}

We assess the uncertainty using $10^5$ simulated trajectories of the model, see also 
Section \ref{sec:uncertainty}.
Figure \ref{fig:M7sims_scen_wUnc} shows the projection results for all the climate state
variables output by Stat-RCM. To visualize how the the different layers of uncertainty impact the projections, Figure \ref{fig:M7sims_TAS_wUnc} contains the first $50$ simulated trajectories in the case of surface temperature, $T^m$. The left panel of Figure \ref{fig:M7sims_TAS_wUnc} includes only uncertainty coming from parameters ($\theta$), while the middle panel additionally includes uncertainty coming  from the innovations in the state equation ($\theta, \eta$), and the right panel adds uncertainty from the measurement errors  ($\theta, \eta, \epsilon$). From the left panel, we learn that uncertainty coming from parameters ($\theta$) is substantial. The resulting simulated trajectories are very smooth, due to the absence of stochastic innovations in these simulations. From the middle panel, we learn that adding uncertainty  from the innovations in the state equation ($\theta, \eta$) increases uncertainty only slightly, which is to be expected since $\widehat{\sigma}_{\eta,m}^2 = Var(\eta_t^m) \approx 0$ (see Table \ref{tab:M7pII} in Appendix \ref{app additional sims}). Conversely, from the right panel, we see that adding uncertainty from the measurement errors  ($\theta, \eta, \epsilon$) increases uncertainty noticeably. These findings indicate that epistemic uncertainty is more important than aleatoric uncertainty for projections of the underlying surface temperature state variable $T_t^m$, while aleatoric uncertainty is crucial in accounting for the variability in  observations of surface temperatures, $T_t^{m,*}$.

%

We may use the model to assess the probability of particular events happening to the
climate system in the future, such as surface temperatures exceeding predefined thresholds,
given a scenario for the covariate processes.
For instance, we might address the question ``What is the probability of exceeding $1.5^\circ$C  at some point in $2023$--$2100$, given some future trajectory of CO$_2$ emissions and other covariate processes?'' We can answer such questions using simulations similar to those above, by reporting the fraction of times the event in question happens in the simulations. Using this methodology for the SSP119 setting discussed in this section, the $1.5^\circ$C threshold is exceeded at some point in the period $2023$--$2100$ with
a $90\%$ probability when considering uncertainty from all parts of the model ($\theta, \eta, \epsilon$).
When considering only uncertainty from uncertainty from parameters ($\theta$) or uncertainty coming from parameters and innovations in the state equation ($\theta, \eta$), however, the probability drops to around $7\%$ and $8\%$, respectively. The $50$ simulated trajectories shown in Figure  \ref{fig:M7sims_TAS_wUnc} illustrate how these numbers come about: When considering only uncertainty from parameters ($\theta$) or from parameters and innovations in the state equation ($\theta, \eta$), the resulting trajectories are reasonably smooth (Figure  \ref{fig:M7sims_TAS_wUnc}, left and middle panels), while if uncertainty from measurement errors are included  ($\theta, \eta, \epsilon$), the resulting trajectories are very volatile (Figure  \ref{fig:M7sims_TAS_wUnc}, right panel). This volatility in the temperature measurements, arising e.g.  from ENSO effects and other natural transitory processes, means that most trajectories will cross the   $1.5^\circ$C threshold (denoted by the horizontal red line in Figure  \ref{fig:M7sims_TAS_wUnc}) at some point in the period $2023$--$2100$. At the same time, the underlying long-term trend, absent the measurement error process $\epsilon$, may plausibly stay below the threshold.

In summary, according to Stat-RCM, there is very high probability  that the $1.5^\circ$C threshold will be breached by temperature observations ($T_t^{m,*}$) at some point over the period  $2023$--$2100$, but, if we filter out the transitory deviations from the underlying temperature trend, represented by $\epsilon$, then there is, in fact, a very good chance that the underlying temperature trend ($T_t^{m}$) will stay below $1.5^\circ$C if the (very ambitious) SSP119 scenario is followed.

%

\newpage
\clearpage


\newpage
\clearpage

\section{Conclusion}\label{sec:concl}

In this study, we have proposed a new statistical reduced-complexity climate model (Stat-RCM). We cast the Stat-RCM in a non-linear state space system that facilitates estimation, filtering, and smoothing using standard statistical methods \citep[][]{durbin2012time}. The model treats anthropogenic CO$_2$ emissions,  forcing from other greenhouse gases,  and natural forcing as covariates.
The climate variables atmospheric concentrations, ocean sink, land sink, forcing from CO$_2$, surface temperature, and deep ocean temperature are modelled
as part of our climate model.

The stochastic formulation of Stat-RCM implies that   tools from statistical theory are available.
A Monte Carlo study has shown that the proposed estimation procedure,
relying on maximizing the log-likelihood function as computed by the extended Kalman filter,
enjoys good finite sample properties. The simulated data for this study have been given similar properties
as those in the historical data record $1959$--$2022$.
When estimation is actually carried out on the historical data,
a statistical model selection procedure has indicated that a forcing equation consisting of a
single logarithmic term is adequate for describing the relationship between
atmospheric concentrations of CO$_2$ and the corresponding radiative forcing.
The resulting model has been validated using two different methods: first, by showing that the estimated
residuals are conform the theoretical expectations; second, by showing that the estimated climate variables
accurately reproduce the historical data over the period 1959--2022.
We regard $(i)$ Monte Carlo simulation studies for assessing the quality of the estimators,
$(ii)$ statistical model selection procedures for informing model specification,
and $(iii)$ statistical validations for checking possible model misspecification, as important features
of empirical statistical modelling.
However, to the best of our knowledge, such statistical methods have not previously
been applied to RCMs that model the entire chain from emissions to temperatures,
due to the physics-based (deterministic) formulation of these models.
Conversely, due to the statistical nature of the Stat-RCM, these analyses can be performed
using well-established methodology.

By extracting the latent temperature state using the extended Kalman filter,
we have estimated the $2022$ long-term global temperature anomaly to be $+1.27^\circ C$
with respect to the $1850$--$1900$ baseline.
We also have used the Stat-RCM to project the climate variables until $2100$,
conditional on a scenario for future CO$_2$ emissions and radiative forcing
from non-CO$_2$ sources. We have found that in the SSP119 scenario, where
emissions are rapidly declining, the $1.5^\circ$C Paris upper limit on the global
surface temperature anomaly, compared to a $1850$--$1900$ baseline, is not a
forgone conclusion. The Stat-RCM estimates a  $90\%$ probability that
the $1.5^\circ$C threshold will be breached by a temperature measurement
$T_t^{m,*}$ in the period $2023$--$2100$ under the SSP119 scenario.
However, if we consider only the underlying temperature trend,
represented by the latent state $T_t^m$, the Stat-RCM estimates
that there is only a $7\%$ probability that the threshold will be breached
in the period $2023$--$2100$ under the SSP119 scenario.
In the latter case, the uncertainty is epistemic
(i.e. stemming from unknown parameters), while, in the former case, aleatoric uncertainty
is also included (i.e. uncertainty from internal variability in the climate system,
including measurement errors and ENSO effects).

We have focused on estimating the Stat-RCM parameters from the historical data
$1959$--$2022$. The main application of RCMs, however, has been as
``emulators'', where the RCM is used to emulate the output of large-scale
climate models \citep[][]{RCMIP2020}. In future work, we intend to employ
the Stat-RCM as an emulator by estimating it using output from large-scale
climate models, such as that those from the Coupled Model Intercomparison
Project \citep[CMIP;][]{Eyring2016}.  This will allow us to employ the
statistical methodology illustrated in this paper to output from CMIP models
with the aim of complementing the results from existing  RCMs,
increasingly used in the IPCC reports, with those from the Stat-RCM.

 {\small 
\vspace{9mm}
\bibliographystyle{chicago}
\bibliography{bhk_references}

\begin{thebibliography}{}

\bibitem[\protect\citeauthoryear{Bacastow and Keeling}{Bacastow and
  Keeling}{1979}]{bacastow1979models}
Bacastow, R.~B. and C.~D. Keeling (1979).
\newblock Models to predict future atmospheric {CO}2 concentrations.
\newblock In {\em Workshop on the global effects of carbon dioxide from fossil
  fuels}, pp.\  72--90. US Department of Energy.

\bibitem[\protect\citeauthoryear{Bennedsen, Hillebrand, and Koopman}{Bennedsen
  et~al.}{2023}]{BHK2023b}
Bennedsen, M., E.~Hillebrand, and S.~J. Koopman (2023).
\newblock A multivariate dynamic statistical model of the global carbon budget
  1959--2020.
\newblock {\em Journal of the Royal Statistical Society Series A: Statistics in
  Society\/}~{\em 186\/}(1), 20--42.

\bibitem[\protect\citeauthoryear{Bennedsen, Hillebrand, and Lykke}{Bennedsen
  et~al.}{2023}]{BHL2022}
Bennedsen, M., E.~Hillebrand, and J.~Z. Lykke (2023).
\newblock Global temperature projections from a statistical energy balance
  model using multiple sources of historical data.
\newblock {\em Journal of Climate\/}~{\em 36\/}(19), 6817--6838.
\newblock Working paper available at \url{https://arxiv.org/abs/2205.10269}.

\bibitem[\protect\citeauthoryear{Betts, Belcher, Hermanson, Tank, Lowe, Jones,
  Morice, Rayner, Scaife, and Stott}{Betts et~al.}{2023}]{Betts2023}
Betts, R.~A., S.~E. Belcher, L.~Hermanson, A.~K. Tank, J.~A. Lowe, C.~D. Jones,
  C.~P. Morice, N.~A. Rayner, A.~A. Scaife, and P.~A. Stott (2023).
\newblock Approaching $1.5^\circ${C}: how will we know we've reached this
  crucial warming mark?
\newblock {\em Nature\/}~{\em 624}, 33--35.

\bibitem[\protect\citeauthoryear{Brockwell and Davis}{Brockwell and
  Davis}{1996}]{BD1996}
Brockwell, P.~J. and R.~A. Davis (1996).
\newblock {\em Introduction to time series and forecasting}.
\newblock Springer New York.

\bibitem[\protect\citeauthoryear{Cheng, Trenberth, Fasullo, Boyer, Abraham, and
  Zhu}{Cheng et~al.}{2017}]{Cheng2017}
Cheng, L., K.~E. Trenberth, J.~Fasullo, T.~Boyer, J.~Abraham, and J.~Zhu
  (2017).
\newblock Improved estimates of ocean heat content from 1960 to 2015.
\newblock {\em Science Advances\/}~{\em 3\/}(3), e1601545.

\bibitem[\protect\citeauthoryear{Cummins, Stephenson, and Stott}{Cummins
  et~al.}{2020}]{Cummins2020}
Cummins, D.~P., D.~B. Stephenson, and P.~A. Stott (2020).
\newblock Optimal estimation of stochastic energy balance model parameters.
\newblock {\em Journal of Climate\/}~{\em 33\/}(18), 7909--7926.

\bibitem[\protect\citeauthoryear{Durbin and Koopman}{Durbin and
  Koopman}{2012}]{durbin2012time}
Durbin, J. and S.~J. Koopman (2012).
\newblock {\em Time series analysis by state space methods}.
\newblock Number~38. Oxford University Press.

\bibitem[\protect\citeauthoryear{Durbin and Watson}{Durbin and
  Watson}{1971}]{Durbin1971}
Durbin, J. and G.~S. Watson (1971).
\newblock Testing for serial correlation in least squares regression.
\newblock {\em Biometrika\/}~{\em 58\/}(1), 1 -- 19.

\bibitem[\protect\citeauthoryear{Engle}{Engle}{1988}]{Engle1988}
Engle, R. (1988).
\newblock Autoregressive conditional heteroscedasticity with estimates of the
  variance of {United Kingdom} inflation.
\newblock {\em Econometrica\/}~{\em 96}, 893--920.

\bibitem[\protect\citeauthoryear{Eyring, Bony, Meehl, Senior, Stevens,
  Stouffer, and Taylor}{Eyring et~al.}{2016}]{Eyring2016}
Eyring, V., S.~Bony, G.~A. Meehl, C.~A. Senior, B.~Stevens, R.~J. Stouffer, and
  K.~E. Taylor (2016).
\newblock {Overview of the Coupled Model Intercomparison Project Phase 6
  (CMIP6) experimental design and organization}.
\newblock {\em Geoscientific Model Development\/}~{\em 9\/}(5), 1937--1958.

\bibitem[\protect\citeauthoryear{Forster, Storelvmo, Armour, Collins, Dufresne,
  Frame, Lunt, Mauritsen, Palmer, Watanabe, Wild, and Zhang}{Forster
  et~al.}{2021}]{ipcc2021c7}
Forster, P., T.~Storelvmo, K.~Armour, W.~Collins, J.-L. Dufresne, D.~Frame,
  D.~Lunt, T.~Mauritsen, M.~Palmer, M.~Watanabe, M.~Wild, and H.~Zhang (2021).
\newblock The {Earth's} energy budget, climate feedbacks, and climate
  sensitivity.
\newblock In V.~Masson-Delmotte, P.~Zhai, A.~Pirani, S.~Connors, C.~P{\'e}an,
  S.~Berger, N.~Caud, Y.~Chen, L.~Goldfarb, M.~Gomis, M.~Huang, K.~Leitzell,
  E.~Lonnoy, J.~Matthews, T.~Maycock, T.~Waterfield, O.~Yelek\c{c}i, R.~Yu, and
  Z.~B. (Eds.), {\em {Climate Change 2021: The Physical Science Basis.
  {C}ontribution of {Working Group I} to the Sixth Assessment Report of the
  {Intergovernmental Panel on Climate Change}}}. Cambridge: Cambridge
  University Press. In Press.

\bibitem[\protect\citeauthoryear{Friedlingstein}{Friedlingstein}{2015}]{Friedlingstein2015}
Friedlingstein, P. (2015).
\newblock Carbon cycle feedbacks and future climate change.
\newblock {\em Philosophical Transactions of the Royal Society A\/}~{\em 373},
  20140421.

\bibitem[\protect\citeauthoryear{Friedlingstein, Dufresne, Cox, and
  Rayner}{Friedlingstein et~al.}{2003}]{Friedlingstein2003}
Friedlingstein, P., J.~L. Dufresne, P.~M. Cox, and P.~Rayner (2003).
\newblock How positive is the feedback between climate change and the carbon
  cycle?
\newblock {\em Tellus B: Chemical and Physical Meteorology\/}~{\em 55\/}(2),
  692--700.

\bibitem[\protect\citeauthoryear{Friedlingstein, O'Sullivan, and
  et~al.}{Friedlingstein et~al.}{2022}]{GCB2022short}
Friedlingstein, P., M.~O'Sullivan, and et~al. (2022).
\newblock {Global Carbon Budget} 2022.
\newblock {\em Earth System Science Data\/}~{\em 14\/}(11), 4811--4900.

\bibitem[\protect\citeauthoryear{Fung, Doney, Lindsay, and John}{Fung
  et~al.}{2005}]{Fung2005}
Fung, I.~Y., S.~C. Doney, K.~Lindsay, and J.~John (2005).
\newblock Evolution of carbon sinks in a changing climate.
\newblock {\em Proceedings of the National Academy of Sciences\/}~{\em
  102\/}(32), 11201--11206.

\bibitem[\protect\citeauthoryear{Gregory}{Gregory}{2000}]{gregory2000vertical}
Gregory, J.~M. (2000).
\newblock Vertical heat transports in the ocean and their effect on
  time-dependent climate change.
\newblock {\em Climate Dynamics\/}~{\em 16\/}(7), 501--515.

\bibitem[\protect\citeauthoryear{Hansen, Fung, Lacis, Rind, Lebedeff, Ruedy,
  Russell, and Stone}{Hansen et~al.}{1988}]{Hansen1988}
Hansen, J., I.~Fung, A.~Lacis, D.~Rind, S.~Lebedeff, R.~Ruedy, G.~Russell, and
  P.~Stone (1988).
\newblock {Global climate changes as forecast by Goddard Institute for Space
  Studies three-dimensional model}.
\newblock {\em Journal of Geophysical Research: Atmospheres\/}~{\em 93\/}(D8),
  9341--9364.

\bibitem[\protect\citeauthoryear{Hansen, Sato, Lacis, Ruedy, Tegen, and
  Matthews}{Hansen et~al.}{1998}]{Hansen1998}
Hansen, J.~E., M.~Sato, A.~Lacis, R.~Ruedy, I.~Tegen, and E.~Matthews (1998).
\newblock Climate forcings in the industrial era.
\newblock {\em Proceedings of the National Academy of Sciences\/}~{\em
  95\/}(22), 12753.

\bibitem[\protect\citeauthoryear{Hartin, Patel, Schwarber, Link, and
  Bond-Lamberty}{Hartin et~al.}{2015}]{HECTOR}
Hartin, C.~A., P.~Patel, A.~Schwarber, R.~P. Link, and B.~P. Bond-Lamberty
  (2015).
\newblock A simple object-oriented and open-source model for scientific and
  policy analyses of the global climate system -- {Hector v1.0}.
\newblock {\em Geoscientific Model Development\/}~{\em 8\/}(4), 939--955.

\bibitem[\protect\citeauthoryear{Haustein, Allen, Forster, Otto, Mitchell,
  Matthews, and Frame}{Haustein et~al.}{2017}]{Haustein:2017aa}
Haustein, K., M.~R. Allen, P.~M. Forster, F.~E.~L. Otto, D.~M. Mitchell, H.~D.
  Matthews, and D.~J. Frame (2017).
\newblock A real-time {Global Warming Index}.
\newblock {\em Scientific Reports\/}~{\em 7\/}(1), 15417.

\bibitem[\protect\citeauthoryear{IPCC}{IPCC}{2023}]{AR6synthesis}
IPCC (2023).
\newblock {Climate Change 2023 Synthesis Report - Summary for policy makers}.
\newblock Technical report, Intergovernmental Panel on Climate Change (IPCC),
  \url{https://www.ipcc.ch/report/ar6/syr/downloads/report/IPCC_AR6_SYR_SPM.pdf}.

\bibitem[\protect\citeauthoryear{Jarque and Bera}{Jarque and
  Bera}{1987}]{JarqueBera1987}
Jarque, C.~M. and A.~K. Bera (1987).
\newblock A test for normality of observations and regression residuals.
\newblock {\em International Statistical Review\/}~{\em 2}, 163--172.

\bibitem[\protect\citeauthoryear{Joos, Bruno, Fink, Siegenthaler, Stocker,
  Qu\'er\'e, and Sarmiento}{Joos et~al.}{1996}]{Joos1996}
Joos, F., M.~Bruno, R.~Fink, U.~Siegenthaler, T.~Stocker, C.~L. Qu\'er\'e, and
  J.~L. Sarmiento (1996).
\newblock An efficient and accurate representation of complex oceanic and
  biospheric models of anthropogenic carbon uptake.
\newblock {\em Tellus\/}~{\em 48B}, 397--417.

\bibitem[\protect\citeauthoryear{Leach, Jenkins, Nicholls, Smith, Lynch, Cain,
  Walsh, Wu, Tsutsui, and Allen}{Leach et~al.}{2021}]{FAIR}
Leach, N.~J., S.~Jenkins, Z.~Nicholls, C.~J. Smith, J.~Lynch, M.~Cain,
  T.~Walsh, B.~Wu, J.~Tsutsui, and M.~R. Allen (2021).
\newblock {FaIR}v2.0.0: a generalized impulse response model for climate
  uncertainty and future scenario exploration.
\newblock {\em Geoscientific Model Development\/}~{\em 14\/}(5), 3007--3036.

\bibitem[\protect\citeauthoryear{Lightfoot and Mamer}{Lightfoot and
  Mamer}{2014}]{LM2014}
Lightfoot, H.~D. and O.~A. Mamer (2014).
\newblock Calculation of atmospheric radiative forcing (warming effect) of
  carbon dioxide at any concentration.
\newblock {\em Energy \& Environment\/}~{\em 25\/}(8), 1439--1454.

\bibitem[\protect\citeauthoryear{Ljung and Box}{Ljung and Box}{1978}]{LB1978}
Ljung, G.~M. and G.~E.~P. Box (1978).
\newblock On a measure of lack of fit in time series models.
\newblock {\em Biometrika\/}~{\em 65\/}(2), 297--303.

\bibitem[\protect\citeauthoryear{Luderer, Vrontisi, Bertram, Edelenbosch,
  Pietzcker, Rogelj, De~Boer, Drouet, Emmerling, Fricko, Fujimori, Havl{\'\i}k,
  Iyer, Keramidas, Kitous, Pehl, Krey, Riahi, Saveyn, Tavoni, Van~Vuuren, and
  Kriegler}{Luderer et~al.}{2018}]{Luderer2018}
Luderer, G., Z.~Vrontisi, C.~Bertram, O.~Y. Edelenbosch, R.~C. Pietzcker,
  J.~Rogelj, H.~S. De~Boer, L.~Drouet, J.~Emmerling, O.~Fricko, S.~Fujimori,
  P.~Havl{\'\i}k, G.~Iyer, K.~Keramidas, A.~Kitous, M.~Pehl, V.~Krey, K.~Riahi,
  B.~Saveyn, M.~Tavoni, D.~P. Van~Vuuren, and E.~Kriegler (2018).
\newblock Residual fossil {CO}2 emissions in 1.5--2$^{\circ}${C} pathways.
\newblock {\em Nature Climate Change\/}~{\em 8\/}(7), 626--633.

\bibitem[\protect\citeauthoryear{Masson-Delmotte and Zhai}{Masson-Delmotte and
  Zhai}{2021}]{IPCC2021_6th}
Masson-Delmotte, V. and P.~Zhai (2021).
\newblock {\em {IPCC Sixth Assessment Report. Climate Change 2021: The Physical
  Science Basis}}.
\newblock Cambridge University Press.

\bibitem[\protect\citeauthoryear{Meinshausen, Meinshausen, Hare, Raper,
  Frieler, Knutti, Frame, and Allen}{Meinshausen
  et~al.}{2009}]{Meinshausen2009}
Meinshausen, M., N.~Meinshausen, W.~Hare, S.~C.~B. Raper, K.~Frieler,
  R.~Knutti, D.~J. Frame, and M.~R. Allen (2009).
\newblock Greenhouse-gas emission targets for limiting global warming to
  2$^\circ${C}.
\newblock {\em Nature\/}~{\em 458\/}(7242), 1158--1162.

\bibitem[\protect\citeauthoryear{Meinshausen, Nicholls, Lewis, Gidden, Vogel,
  Freund, Beyerle, Gessner, Nauels, Bauer, Canadell, Daniel, John, Krummel,
  Luderer, Meinshausen, Montzka, Rayner, Reimann, Smith, van~den Berg, Velders,
  Vollmer, and Wang}{Meinshausen et~al.}{2020}]{SSP2020}
Meinshausen, M., Z.~R.~J. Nicholls, J.~Lewis, M.~J. Gidden, E.~Vogel,
  M.~Freund, U.~Beyerle, C.~Gessner, A.~Nauels, N.~Bauer, J.~G. Canadell, J.~S.
  Daniel, A.~John, P.~B. Krummel, G.~Luderer, N.~Meinshausen, S.~A. Montzka,
  P.~J. Rayner, S.~Reimann, S.~J. Smith, M.~van~den Berg, G.~J.~M. Velders,
  M.~K. Vollmer, and R.~H.~J. Wang (2020).
\newblock {The shared socio-economic pathway (SSP) greenhouse gas
  concentrations and their extensions to 2500}.
\newblock {\em Geoscientific Model Development\/}~{\em 13\/}(8), 3571--3605.

\bibitem[\protect\citeauthoryear{Meinshausen, Raper, and Wigley}{Meinshausen
  et~al.}{2011}]{MAGICC}
Meinshausen, M., S.~C.~B. Raper, and T.~M.~L. Wigley (2011).
\newblock Emulating coupled atmosphere-ocean and carbon cycle models with a
  simpler model, {MAGICC}6 -- {Part} 1: {Model} description and calibration.
\newblock {\em Atmospheric Chemistry and Physics\/}~{\em 11\/}(4), 1417--1456.

\bibitem[\protect\citeauthoryear{Meinshausen, Smith, Calvin, Daniel, Kainuma,
  Lamarque, Matsumoto, Montzka, Raper, Riahi, Thomson, Velders, and van
  Vuuren}{Meinshausen et~al.}{2011}]{Meinshausen2011}
Meinshausen, M., S.~J. Smith, K.~Calvin, J.~S. Daniel, M.~L.~T. Kainuma, J.-F.
  Lamarque, K.~Matsumoto, S.~A. Montzka, S.~C.~B. Raper, K.~Riahi, A.~Thomson,
  G.~J.~M. Velders, and D.~P.~P. van Vuuren (2011).
\newblock {The RCP greenhouse gas concentrations and their extensions from 1765
  to 2300}.
\newblock {\em Climatic Change\/}~{\em 109\/}(1), 213.

\bibitem[\protect\citeauthoryear{Morice, Kennedy, Rayner, Winn, Hogan, Killick,
  Dunn, Osborn, Jones, and Simpson}{Morice et~al.}{2020}]{morice2020updated}
Morice, C.~P., J.~J. Kennedy, N.~A. Rayner, J.~Winn, E.~Hogan, R.~Killick,
  R.~Dunn, T.~Osborn, P.~Jones, and I.~Simpson (2020).
\newblock \href{https://doi.org/10.1029/2019JD032361}{An updated assessment of
  near-surface temperature change from 1850: The HadCRUT5 dataset}.
\newblock {\em Journal of Geophysical Research: Atmospheres\/}.

\bibitem[\protect\citeauthoryear{Nicholls, Meinshausen, and et~al.}{Nicholls
  et~al.}{2021}]{RCMIP2021}
Nicholls, Z., M.~Meinshausen, and et~al. (2021).
\newblock Reduced complexity model intercomparison project phase 2:
  Synthesizing earth system knowledge for probabilistic climate projections.
\newblock {\em Earth's Future\/}~{\em 9\/}(6), 1--25.

\bibitem[\protect\citeauthoryear{Nicholls, Meinshausen, Lewis, Gieseke,
  Dommenget, Dorheim, Fan, Fuglestvedt, Gasser, Gol\"uke, Goodwin, Hartin,
  Hope, Kriegler, Leach, Marchegiani, McBride, Quilcaille, Rogelj, Salawitch,
  Samset, Sandstad, Shiklomanov, Skeie, Smith, Smith, Tanaka, Tsutsui, and
  Xie}{Nicholls et~al.}{2020}]{RCMIP2020}
Nicholls, Z. R.~J., M.~Meinshausen, J.~Lewis, R.~Gieseke, D.~Dommenget,
  K.~Dorheim, C.-S. Fan, J.~S. Fuglestvedt, T.~Gasser, U.~Gol\"uke, P.~Goodwin,
  C.~Hartin, A.~P. Hope, E.~Kriegler, N.~J. Leach, D.~Marchegiani, L.~A.
  McBride, Y.~Quilcaille, J.~Rogelj, R.~J. Salawitch, B.~H. Samset,
  M.~Sandstad, A.~N. Shiklomanov, R.~B. Skeie, C.~J. Smith, S.~Smith,
  K.~Tanaka, J.~Tsutsui, and Z.~Xie (2020).
\newblock {Reduced Complexity Model Intercomparison Project Phase 1:
  introduction and evaluation of global-mean temperature response}.
\newblock {\em Geoscientific Model Development\/}~{\em 13\/}(11), 5175--5190.

\bibitem[\protect\citeauthoryear{Pretis}{Pretis}{2020}]{Pretis2020}
Pretis, F. (2020).
\newblock Econometric modelling of climate systems: {T}he equivalence of energy
  balance models and cointegrated vector autoregressions.
\newblock {\em Journal of Econometrics\/}~{\em 214\/}(1), 256--273.

\bibitem[\protect\citeauthoryear{Ramaswami}{Ramaswami}{2001}]{Ramaswami2001}
Ramaswami, V. (2001).
\newblock Chapter 6: Radiative forcing of climate change.
\newblock Technical report, IPCC.

\bibitem[\protect\citeauthoryear{Sanderson, O'Neill, and Tebaldi}{Sanderson
  et~al.}{2016}]{SONBT2016}
Sanderson, B.~M., B.~C. O'Neill, and C.~Tebaldi (2016).
\newblock What would it take to achieve the {Paris} temperature targets?
\newblock {\em Geophysical Research Letters\/}~{\em 43\/}(13), 7133--7142.

\bibitem[\protect\citeauthoryear{Schwartz}{Schwartz}{2007}]{Schwartz2007}
Schwartz, S.~E. (2007).
\newblock Heat capacity, time constant, and sensitivity of {E}arth's climate
  system.
\newblock {\em Journal of Geophysical Research\/}~{\em 112\/}(D24), 1--12.

\bibitem[\protect\citeauthoryear{Schwarz}{Schwarz}{1978}]{BIC1978}
Schwarz, G. (1978).
\newblock Estimating the dimension of a model.
\newblock {\em The Annals of Statistics\/}~{\em 6\/}(2), 461--464.

\bibitem[\protect\citeauthoryear{Shi}{Shi}{1992}]{Shi1992}
Shi, G. (1992).
\newblock Radiative forcing and greenhouse effect due to the atmospheric trace
  gases.
\newblock {\em Science in China (Series B)\/}~{\em 35}, 217--229.

\bibitem[\protect\citeauthoryear{Smith, Cummins, Fredriksen, Nicholls,
  Meinshausen, Allen, Jenkins, Leach, Mathison, and Partanen}{Smith
  et~al.}{2024}]{FAIR_calibrate}
Smith, C., D.~P. Cummins, H.-B. Fredriksen, Z.~Nicholls, M.~Meinshausen,
  M.~Allen, S.~Jenkins, N.~Leach, C.~Mathison, and A.-I. Partanen (2024).
\newblock fair-calibrate v1.4.1: calibration, constraining and validation of
  the fair simple climate model for reliable future climate projections.
\newblock {\em EGUsphere\/}, 1--36.

\bibitem[\protect\citeauthoryear{Smith, Nicholls, Armour, Collins, Forster,
  Meinshausen, Palmer, and Watanabe}{Smith et~al.}{2021}]{ipcc2021supp7}
Smith, C., Z.~Nicholls, K.~Armour, W.~Collins, P.~Forster, M.~Meinshausen,
  M.~Palmer, and M.~Watanabe (2021).
\newblock The earth's energy budget, climate feedbacks, and climate sensitivity
  supplementary material.
\newblock In V.~Masson-Delmotte, P.~Zhai, A.~Pirani, S.~Connors, C.~P{\'e}an,
  S.~Berger, N.~Caud, Y.~Chen, L.~Goldfarb, M.~Gomis, M.~Huang, K.~Leitzell,
  E.~Lonnoy, J.~Matthews, T.~Maycock, T.~Waterfield, O.~Yelek\c{c}i, R.~Yu, and
  Z.~B. (Eds.), {\em {Climate Change 2021: The Physical Science Basis.
  {C}ontribution of {Working Group I} to the Sixth Assessment Report of the
  {Intergovernmental Panel on Climate Change}}}. Cambridge: Cambridge
  University Press. In Press.

\bibitem[\protect\citeauthoryear{Solomon, Qin, and Manning}{Solomon
  et~al.}{2007}]{IPCC2007_4th}
Solomon, S., D.~Qin, and M.~Manning (2007).
\newblock {\em {IPCC Fourth Assessment Report. Climate Change 2007: The
  Physical Science Basis}}.
\newblock Cambridge University Press.

\bibitem[\protect\citeauthoryear{Stocker and Qin}{Stocker and
  Qin}{2013}]{IPCC2013_5th}
Stocker, T.~F. and D.~Qin (2013).
\newblock {\em {IPCC Fifth Assessment Report. Climate Change 2013: The Physical
  Science Basis}}.
\newblock Cambridge University Press.

\bibitem[\protect\citeauthoryear{Tokarska and Gillett}{Tokarska and
  Gillett}{2018}]{TG2018}
Tokarska, K.~B. and N.~P. Gillett (2018).
\newblock Cumulative carbon emissions budgets consistent with 1.5
  $\,^{\circ}${C} global warming.
\newblock {\em Nature Climate Change\/}~{\em 8\/}(4), 296--299.

\bibitem[\protect\citeauthoryear{UNFCCC}{UNFCCC}{2015}]{FCCC2015}
UNFCCC (2015).
\newblock Adoption of the {Paris} {A}greement,
  \url{https://unfccc.int/resource/docs/2015/cop21/eng/l09r01.pdf}.
\newblock Technical report, United Nations Framework Convention on Climate
  Change.

\bibitem[\protect\citeauthoryear{Zhang, Wang, Rayner, Ciais, Huang, Luo, Piao,
  Wang, Xia, Zhao, Zheng, Tian, and Zhang}{Zhang et~al.}{2021}]{ZhangX2021}
Zhang, X., Y.-P. Wang, P.~Rayner, P.~Ciais, K.~Huang, Y.~Luo, S.~Piao, Z.~Wang,
  J.~Xia, W.~Zhao, X.~Zheng, J.~Tian, and Y.~Zhang (2021).
\newblock A small climate-amplifying effect of climate-carbon cycle feedback.
\newblock {\em Nature Communications\/}~{\em 12}, 2952.

\end{thebibliography}
}

\appendix

\section{Details of the state space representation of Stat-RCM} \label{app:ssr}
Let $y_t = (C_t^*,S_t^{OCN,*},S_t^{LND,*},F_t^{CO2,*},T_t^{m,*},T_t^{d,*},O_t^*)'$ denote the  $7 \times 1$ vector of observations at time $t$ and $\Delta>0$ the length of a time step between observations. In Section  \ref{sec:system} we presented the non-linear state space model of the climate system, given as follows
\[
y_t          \ = \ \mu + A x_t + \epsilon_t, \qquad  \qquad
x_{t+\Delta}   \ = \ B(x_t) + W_t +  R \eta_{t,\Delta},
\]
where $x_t = (C_t, S_t^{OCN}, S_t^{LND}, F_t^{CO2}, T_t^{m}, T_t^{d})'$ denotes the $6 \times 1$ latent state vector. At time $t$, there are thus $7$ observations, collected in $y_t$, for $6$ states, collected in $x_t$. The  $7 \times 1$  vector $\mu = (\mu_C, \mu_L, \mu_O, \mu_F, \mu_m, \mu_d, H_d \cdot  \mu_d)'$  contains the intercepts in the measurement equations, and the  $7 \times 6$  matrix $A$ captures the relations between the observations $y_t$ and the underlying state vector $x_t$ described in Section \ref{sec:measurement},
\begin{align*}
A  = \begin{pmatrix}
    1 & 0 & 0 & 0 & 0 & 0 \\
    0 & 1 & 0 & 0 & 0 & 0 \\
    0 & 0 & 1 & 0 & 0 & 0 \\
    0 & 0 & 0 & 1 & 0 & 0 \\
    0 & 0 & 0 & 0 & 1 & 0 \\
    0 & 0 & 0 & 0 & 0 & 1 \\
    0 & 0 & 0 & 0 & 0 & H_d 
    \end{pmatrix}.
\end{align*}
The entries in the  $7 \times 1$  measurement error vector $\epsilon_t$ are modelled as AR(1) processes,
$$\epsilon_{t+\Delta} = \Phi \epsilon_{t} + \xi_{t,\Delta},$$
with $\Phi$ being a  $7 \times 7$ diagonal matrix given by $\Phi  = \text{diag}(\phi_1,\phi_2, \ldots, \phi_7)$,
with $\phi_i \in (-1,1)$ for all $i$, and $\xi_{t,\Delta} \stackrel{iid}{\sim} N(0,\Delta \cdot P)$
with $P$ being the $7 \times 7$ matrix
\begin{align*}
P  = \begin{pmatrix}
    \sigma^2_{\epsilon,C} & 0 & 0 & 0 & 0 & 0 & 0 \\
    0 & \sigma^2_{\epsilon,OCN} & 0 & 0 & 0 & 0 & 0 \\
    0 & 0 &  \sigma^2_{\epsilon,LND} & 0 & 0 & 0 & 0 \\
    0 & 0 & 0 &  \sigma^2_{\epsilon,F} & 0 & 0 & 0 \\
    0 & 0 & 0 & 0 &  \sigma^2_{\epsilon,m} & 0 & 0 \\
    0 & 0 & 0 & 0 & 0 &  \sigma^2_{\epsilon,d} & \rho \sigma_{\epsilon,d} \sigma_{\epsilon,OHC} \\
    0 & 0 & 0 & 0 & 0 & \rho  \sigma_{\epsilon,d} \sigma_{\epsilon,OHC} & \sigma_{\epsilon,OHC}^2
    \end{pmatrix},
\end{align*}
where $\rho \in (-1,1)$ allows for correlation in the measurement errors of
deep ocean temperature $T_t^{d,*}$ and ocean heat content $O_t^*$, see \cite{BHL2022}.

The non-linear  $6 \times 1$  mapping $B(\cdot)$ represents the state transition equations as described in Section \ref{sec:ss},
\begin{align*}
B(x_t)  = \begin{pmatrix}
    C_t \\
     b_1 C_t \exp(-c_1 T_t^d) \\
      b_2 C_t \exp(-c_2 T_t^d) \\
      f_1 \log (C_t + f_2 C_t^2) + f_3 \sqrt{C_t} \\
      \left(1-\frac{\gamma+\lambda}{H_m} \Delta \right)T^m_t + \frac{\gamma \Delta}{H_m} T^d_{t}  \\
       \frac{\gamma \Delta}{H_d}   T^m_t + \left( 1 - \frac{\gamma \Delta}{H_d} \right)T^d_t
    \end{pmatrix},
\end{align*}
while the  $6 \times 1$  vector $W_t$ collects constants and covariates into the state equation,
\begin{align*}
W_t  = \begin{pmatrix}
    C_t -\Delta(b_1+b_2)C_{1750}   + \Delta E_{t+\Delta}  \\
     -b_1 C_{1750}  \\
      -b_2 C_{1750}  \\
     -   f_1 \log ( C_{1750}  + f_2  C_{1750}^2) + f_3 \sqrt{ C_{1750} } \\
     \frac{\Delta}{H_m}(F_t^{CO2} + F_t^{Ex})   \\
      0
    \end{pmatrix}.
\end{align*}
The $5 \times 1$  innovation sequence $\eta_ {t,\Delta}  \stackrel{iid}{\sim} N(0,\Delta \cdot Q)$ is modelled with a  $5 \times 5$  diagonal covariance matrix $Q = \text{diag}(\sigma_{\eta,OCN}^2,  \sigma_{\eta,LND}^2, \sigma_{\eta,F}^2,\sigma_{\eta,m}^2, \sigma_{\eta,d}^2)$ and the  $6 \times 5$  matrix $R$ completes the specification for the state variables, 
 \begin{align*}
R = \begin{pmatrix}
    -\Delta & - \Delta & 0 & 0 & 0 \\
    1 & 0 & 0 & 0 & 0 \\
    0 & 1 & 0 & 0 & 0 \\
    0 & 0 & 1 & 0 & 0 \\
    0 & 0 & 0 & 1 & 0 \\
    0 & 0 & 0 & 0 & 1
    \end{pmatrix}.
\end{align*}

Letting $\phi = (\phi_1,\ldots, \phi_7)'$, $\sigma_\epsilon^2 = (\sigma_{\epsilon,C}^2,\sigma_{\epsilon,OCN}^2,\sigma_{\epsilon,LND}^2,\sigma_{\epsilon,F}^2,\sigma_{\epsilon,m}^2,\sigma_{\epsilon,d}^2,\sigma_{\epsilon,O}^2)'$, $\sigma_{\eta}^2 = (\sigma_{\eta,OCN}^2, \sigma_{\eta,LND}^2,\sigma_{\eta,F}^2,\sigma_{\eta,m}^2, \sigma_{\eta,d}^2)$ we arrive at the $37 \times 1$ vector $\theta$ of unknown parameters, given by
 \begin{align*}
\theta = (b_1,b_2,c_1,c_2,f_1,f_2,f_3,\gamma,\lambda,H_m,H_d,\mu_C, \mu_L, \mu_O, \mu_F, \mu_m, \mu_d,\rho, \phi', \sigma_\epsilon',\sigma_\eta')'.
\end{align*}
Note that we in our implementation of the Stat-RCM impose several restrictions on the elements in $\theta$, so that the resulting dimension of the parameter vector is smaller than $37$. Our preferred specification of the Stat-RCM sets $\mu_C = \mu_L = \mu_O = \mu_F = 0$ and $f_2 = f_3 = 0$, resulting in $\theta$ being a $31 \times 1$ vector, see Section \ref{sec:empirical} of the main paper.

In our implementation, we estimate the parameter vector $\theta$ by the method of maximum likelihood, using the extended Kalman filter \citep[][Chapter 10]{durbin2012time}. This procedure requires the Jacobian of the transition function $B(\cdot)$, which is given by
\begin{align*}
\frac{\partial B(x_t)}{\partial x_t'}  = \begin{pmatrix}
    1  & 0 & 0 & 0 & 0  & 0 \\
     b_1 \exp(-c_1 T_t^d) & 0 & 0 & 0 & 0  &  - b_1 c_1 C_t \exp(c_1 T_t^d)  \\
       b_2 \exp(-c_2 T_t^d) & 0 & 0 & 0 & 0  &  - b_2 c_2 C_t \exp(c_2 T_t^d)  \\
       \frac{f_1}{ C_t + f_2 C_t^2} \left( 1+ 2f_2 C_t\right) + \frac{f_3}{2 \sqrt{C_t}}  & 0 & 0 & 0 & 0  & 0   \\
     0  & 0 & 0 & 0 & \left(1-\frac{\gamma+\lambda}{H_m} \Delta \right)  &  \frac{\gamma \Delta}{H_m}  \\
     0  & 0 & 0 & 0 &  \frac{\gamma \Delta}{H_d}    &  \left( 1 - \frac{\gamma \Delta}{H_d} \right)
    \end{pmatrix}.
\end{align*}

\section{Additional parameter estimates from the Stat-RCM} \label{app additional sims}

Section \ref{sec estimation results} of the main paper presents the estimates of the physical parameters of the Stat-RCM when applied to the historical data over the period $1959$--$2019$. The estimated parameters relating to the state ($\eta_\cdot$) and measurement ($\epsilon_\cdot$) error processes are shown in Tables \ref{tab:M7pII}--\ref{tab:M7pIII}.  We see that some estimates of the variances of the state disturbances $\eta_\cdot$ are estimated to be close to zero, in particular $ \widehat\sigma_{\eta,F}^2,\widehat\sigma_{\eta,m}^2 \approx 0$. This indicates that for these state variables -- corresponding to forcing from CO2, $ F_t^{CO2}$, and surface temperature, $T_t^m$ -- the internal state dynamics as specified by the three modules of the Stat-RCM are sufficient to capture the evolution of the states. For the remaining state variables, we have $\widehat \sigma_{\eta,OCN}^2, \widehat \sigma_{\eta,LND}^2, \widehat \sigma_{\eta,d}^2>0$, indicating that for these states, an additional disturbance term must be included to capture the evolution of the states of the sinks $S_t^{OCN}$ and $S_t^{LND}$ and the deep ocean temperature $T_t^d$.

It is also worth remarking that $\widehat \rho = 0.87$, highlighting the close connection between the data on deep ocean temperature ($T_t^{d,*}$) and ocean heat content ($O_t^{*}$). We refer to  \cite{BHL2022} for a more in-depth discussion on the relation between these two time series.

%


\begin{table}
\caption{\it Stat-RCM parameter estimates II}
\begin{center}
\tiny
\begin{tabularx}{1.00\textwidth}{@{\extracolsep{\stretch{1}}}l|ccccccccccccc@{}} 
\toprule
 & $\sigma_{\eta,OCN}^2$  & $  \sigma_{\eta,LND}^2$  & $\sigma_{\eta,F}^2$  & $\sigma_{\eta,m}^2$  & $ \sigma_{\eta,d}^2$ & $\sigma_{\epsilon,C}^2$ & $\sigma_{\epsilon,OCN}^2$ & $\sigma_{\epsilon,LND}^2$ & $\sigma_{\epsilon,F}^2$ & $\sigma_{\epsilon,m}^2$ & $\sigma_{\epsilon,d}^2$  & $\sigma_{\epsilon,O}^2$  & $\rho$ \\  
\midrule
Estimate:  & $   0.00 $ & $   0.70 $ & $   0.00 $ & $   0.00 $ & $   0.01 $ & $   0.45 $ & $   0.12 $ & $   0.42 $ & $   0.01 $ & $   0.09 $ & $   0.00 $ & $   0.04 $ & $   0.87 $\\
Std. Err.:  & $(   0.03 )$ & $(   0.08 )$ & $(   0.00 )$ & $(   0.02 )$ & $(   0.00 )$ & $(   0.06 )$ & $(   0.01 )$ & $(   0.07 )$ & $(   0.00 )$ & $(   0.01 )$ & $(   0.00 )$ & $(   0.02 )$ & $(   0.21 )$\\
$t$-stat:  & $   0.03 $ & $   9.17 $ & $   0.00 $ & $   0.01 $ & $  11.11 $ & $   7.20 $ & $  11.01 $ & $   5.73 $ & $   8.97 $ & $  11.07 $ & $   1.18 $ & $   2.71 $ & $   4.16 $\\
\bottomrule 
\end{tabularx}
\end{center}
{\footnotesize \it Parameter estimates by the maximum likelihood method applied to the $1959$--$2022$ data.} 
\label{tab:M7pII}
\end{table}

\begin{table}
\caption{\it Stat-RCM: Parameter estimates III}
\begin{center}
\tiny
\begin{tabularx}{1.00\textwidth}{@{\extracolsep{\stretch{1}}}l|ccccccc@{}} 
\toprule
 & $\phi_1$ & $\phi_2$  & $\phi_3$ & $\phi_4$ & $\phi_5$ &  $\phi_6$ &  $\phi_7$  \\  
\midrule
Estimate:  & $   0.78 $ & $   0.56 $ & $   0.39 $ & $   0.57 $ & $   0.15 $ & $   0.89 $ & $   0.89 $\\
Std. Err.:  & $(   0.08 )$ & $(   0.11 )$ & $(   0.21 )$ & $(   0.15 )$ & $(   0.13 )$ & $(   0.07 )$ & $(   0.07 )$\\
$t$-stat (Asym):  & $  10.00 $ & $   4.92 $ & $   1.90 $ & $   3.80 $ & $   1.16 $ & $  12.12 $ & $  12.11 $\\
\bottomrule 
\end{tabularx}
\end{center}
{\footnotesize \it Parameter estimates by the maximum likelihood method applied to the $1959$--$2022$ data.} 
\label{tab:M7pIII}
\end{table}

%
%
%
%
%

%
%
%

\end{document}